\documentclass[preprint]{iucr}              

     \journalcode{S}              

\newlength{\figurewidth}
\newlength{\figuredoublewidth}

\usepackage{amsmath}
\usepackage{xcolor}
\usepackage{siunitx}
\usepackage[colorinlistoftodos]{todonotes}
\usepackage{multirow}

\begin{document}                  




\title{Photon shot-noise limited
transient absorption soft X-ray spectroscopy at the European XFEL}
\shorttitle{Transient XAS at SCS}




\cauthor[a]{Lo\"ic}{Le Guyader}{loic.le.guyader@xfel.eu}

\author[b]{Andrea}{Eschenlohr} 
\author[c]{Martin}{Beye} 
\author[d]{William}{Schlotter} 
\author[e]{Florian}{D\"oring}
\author[a]{Cammille}{Carinan} 
\author[a]{David}{Hickin} 

\aff[a]{European XFEL, Holzkoppel 4, 22869 Schenefeld, Germany}
\aff[b]{Faculty of Physics and Center for Nanointegration Duisburg-Essen (CENIDE), University Duisburg-Essen, Lotharstr. 1, 47057 Duisburg, Germany}
\aff[c]{Deutsches Elektronen-Synchrotron DESY, Notkestr. 85, 22607 Hamburg, German}
\aff[d]{Linear Coherent Light Source, SLAC National Accelerator Lab, 2575 Sand Hill Rd., Menlo Park, CA 94025, USA}
\aff[e]{Paul Scherrer Institute, 5232 Villigen PSI, Switzerland}

\author[a]{Naman}{Agarwal}

\author[f]{Christine}{Boeglin}
\aff[f]{Université de Strasbourg, CNRS, Institut de Physique et Chimie des Matériaux de Strasbourg, UMR 7504, Strasbourg, F-67000 France}
\aff[g]{Department of Physics and Center for Applied Photonics, University of Konstanz, D-78457 Konstanz, Germany}
\aff[h]{Department of Physics, TU Dortmund University, Otto-Hahn Straße 4, 44227 Dortmund, Germany}
\aff[i]{Department of Physics, Stockholm University, 106 91 Stockholm, Sweden}
\aff[j]{Department of Molecular Sciences and Nanosystems, Ca’ Foscari University of Venice, 30172 Venice, Italy}
\author[b]{Uwe}{Bovensiepen}
\author[c]{Jens}{Buck}

\author[a]{Robert}{Carley}
\author[k]{Andrea}{Castoldi} 
\aff[k]{Politecnico di Milano, Dip. Elettronica, Informazione e Bioingegneria and INFN, Sezione di Milano, Milano, Italy}

\author[l]{Alessandro}{D'Elia}
\aff[l]{IOM-CNR, Laboratorio Nazionale TASC, Basovizza SS-14, km 163.5, 34012 Trieste, Italy}
\author[a]{Jan-Torben}{Delitz}

\author[a]{Wajid}{Ehsan}
\author[c]{Robin}{Engel}
\author[o]{Florian}{Erdinger} 
\aff[o]{Institute for Computer Engineering, University of Heidelberg, Mannheim, Germany}

\author[a,o2,o3]{Hans}{Fangohr} 
\aff[o2]{Max-Planck Institute for the Structure and Dynamics of Matter, Luruper Chaussee 149, 22761 Hamburg, Germany}
\aff[o3]{University of Southampton, Southampton SO17 1BJ, United Kingdom}
\author[o]{Peter}{Fischer} 
\author[k]{Carlo}{Fiorini} 
\author[p]{Alexander}{F\"ohlisch} 
\aff[p]{Institute for Methods and Instrumentation for Synchrotron Radiation Research (PS-ISRR),
Helmholtz-Zentrum Berlin für Materialien und Energie GmbH (HZB), Albert-Einstein Straße 15, 12489 Berlin, Germany}

\author[a]{Luca}{Gelisio}
\author[q,r]{Michael}{Gensch}
\aff[q]{Institute of Optical Sensor Systems, DLR (German Aerospace Center), Rutherfordstr. 2, 12489 Berlin, Germany}
\aff[r]{Institute of Optics and Atomic Physics, Technische Universität Berlin, Strasse des 17. Juni 135, 10623 Berlin, Germany}
\author[a]{Natalia}{Gerasimova}
\author[a]{Rafael}{Gort}

\author[c]{Karsten}{Hansen} 
\author[a]{Steffen}{Hauf}

\author[a]{Manuel}{Izquierdo}

\author[u]{Emmanuelle}{Jal}

\author[a]{Ebad}{Kamil}
\author[a]{Suren}{Karabekyan}
\aff[v]{Molecular Biophysics and Integrated Bioimaging Division, Lawrence Berkeley National Laboratory, Berkeley, CA, USA}
\author[a]{Thomas}{Kluyver}

\author[c,w]{Tim}{Laarmann}
\aff[w]{The Hamburg Centre for Ultrafast Imaging CUI, Luruper Chaussee 149, 22761 Hamburg, Germany}
\author[b]{Tobias}{Lojewski}
\author[a]{David}{Lomidze}

\author[c]{Stefano}{Maffessanti} 
\author[e]{Talgat}{Mamyrbayev}
\author[y,z,0]{Augusto}{Marcelli}
\aff[y]{INFN – Laboratori Nazionali di Frascati, via Enrico Fermi 54, 00044 Frascati, Italy}
\aff[z]{RICMASS – Rome International Center for Materials Science Superstripes, 00185 Rome, Italy}
\aff[0]{Istituto Struttura della Materia, CNR, Via del Fosso del Cavaliere 100, 00133 Rome, Italy}
\author[a]{Laurent}{Mercadier}
\author[a]{Giuseppe}{Mercurio}
\author[a,c]{Piter S.}{Miedema}


\author[b]{Katharina}{Ollefs}

\author[c1,c2]{Kai}{Rossnagel}
\aff[c1]{Institute of Experimental and Applied Physics, Kiel University, 24098 Kiel, Germany}
\aff[c2]{Ruprecht Haensel Laboratory, Deutsches Elektronen-Synchrotron DESY, 22607 Hamburg, Germany}
\author[e]{Benedikt}{R\"osner}
\author[b]{Nico}{Rothenbach}

\author[a]{Andrey}{Samartsev}
\author[a]{Justine}{Schlappa}
\author[a]{Kiana}{Setoodehnia}
\author[u]{Gheorghe}{Sorin Chiuzbaian}
\author[b]{Lea}{Spieker} 
\author[3]{Christian}{Stamm}
\aff[3]{Department of Materials, ETH Zürich, 8093 Zürich, Switzerland}
\author[4]{Francesco}{Stellato}
\aff[4]{Physics Department, University of Rome Tor Vergata, and INFN—Sezione di Roma Tor Vergata, Via della Ricerca Scientifica 1, 00133 Roma, Italy}

\author[c]{Simone}{Techert}
\author[a]{Martin}{Teichmann}
\author[a]{Monica}{Turcato}

\author[a]{Benjamin}{Van Kuiken} 

\author[b]{Heiko}{Wende}
\author[a]{Alexander}{Yaroslavtsev}

\author[a]{Jun}{Zhu} 


\author[a]{Serguei}{Molodtsov}
\author[e]{Christian}{David} 
\author[a,j]{Matteo}{Porro}
\author[a]{Andreas}{Scherz}









\newcommand{\ts}{\textsuperscript}

\maketitle                        

\begin{synopsis}
A beam-splitting off-axis zone plate setup to measure transient X-ray Absorption
Spectroscopy (XAS) is presented, as implemented at the Spectroscopy \& Coherent
Scattering (SCS) instrument at the European X-ray Free Electron Laser.
\end{synopsis}

\begin{abstract}
Femtosecond transient soft X-ray Absorption Spectroscopy (XAS) is a very
promising technique that can be employed at X-ray Free Electron Lasers (FELs)
to investigate out-of-equilibrium dynamics for material and energy research. Here we
present a dedicated setup for soft X-rays available at the Spectroscopy \&
Coherent Scattering (SCS) instrument at the European X-ray Free Electron Laser
(EuXFEL). It consists of a beam-splitting off-axis zone plate (BOZ) used in
transmission to create three copies of the incoming beam, which are used to
measure the transmitted intensity through the excited and unexcited sample, as
well as to monitor the incoming intensity.
Since these three intensity signals are detected shot-by-shot and simultaneously,
this setup allows normalized shot-by-shot analysis of the transmission. For photon
detection, the DSSC imaging detector, which is capable of recording up to 800 images
at 4.5~MHz
frame rate during the FEL burst, is employed and allows approaching the photon
shot-noise limit. We review the setup and its capabilities, as well as the online and
offline analysis tools provided to users.
\end{abstract}


\section{Introduction}

%

X-ray Absorption Spectroscopy is one of the most widely used techniques at
synchrotron radiation facilities around the world, aimed at the investigation of the local structure and
electronic properties of atoms in solids and molecules on surfaces or in
solutions \cite{Bianconi1992, stohr1992nexafs, Bokhoven2016}. Its implementation at FELs opens the possibility
of performing high-resolution spectroscopy like at synchrotrons, with the
added advantage of accessing ultrafast dynamics on the
femtosecond timescale. Transient XAS allows, for example, monitoring 
electron-hole dynamics \cite{Boeglin2010, Zuerch2017, Britz2021}, electron localization \cite{Stamm2007, Lojewski2022},
on-site Coulomb repulsion \cite{Baykusheva2022}, lattice excitation \cite{Rothenbach2019, Rothenbach2021}, magnetic order \cite{Agarwal2022} and
ultrafast phase transition \cite{Cavalleri2005}. Monochromatic soft
X-ray pulses of few nJ of energy are easily delivered by FELs and contain few
10$^7$ photons. At the photon shot-noise limit, a single shot signal-to-noise ratio
(SNR) at the level of a few thousands is thus achievable. However, the
X-ray pulses generated by self-amplified spontaneous emission (SASE)
at an FEL feature very high pulse-to-pulse intensity
fluctuations after a monochromator \cite{Saldin1998}. It is therefore essential
to normalize the
transmitted signal by measuring the incoming radiation intensity before the
sample ($I_\mathrm{o}$). The main challenge in measuring femtosecond XAS in the soft X-ray
regime, where small changes in the spectra have to be detected, is precisely this
normalization scheme.

XAS at FELs was pioneered by \cite{Bernstein2009}, where the $I_\mathrm{o}$
normalization was achieved by using a half of a sample, such that one half of the X-ray
beam was propagating through the sample, while the other half was propagating freely.
A Ce-doped yttrium aluminum garnet (Ce:YAG) scintillator screen placed in front
of an intensified charge coupled device camera (ICCD) was used for detection.
This approach relies on spatial beam coherence and pointing stability of the FEL beam.
An alternative approach consists of using a transmission grating to create copies of
the incoming beam with the different diffraction orders, and using the +1\ts{st}
grating order to measure the sample transmission and the -1\ts{st} grating
order to measure the $I_\mathrm{o}$ \cite{Katayama2013, Katayama2016, Brenner2019,
Engel2020, Engel2021}. The advantage here is that the beam intensities are linked
by the grating element.
Later, the sensitivity of this method was improved by combining the transmission-grating approach
with a focusing zone plate component. As the beams propagate towards the detector,
they are focused in front of the sample and then expand and illuminate many more pixels,
thereby increasing the maximum number
of photons that can be detected without saturating the detector \cite{Schlotter2020}.
In our improved scheme, we use an off-axis zone plate \cite{Buzzi2017, Jal2019, Roesner2020}
which gives the possibility of separating the different zone plate orders on the detector, as
we shall discuss.

In this article, we will review the scheme as implemented at the SCS instrument at the
European XFEL. In sec.~\ref{sec:setup}, the setup is described. First, an overview of the
setup and its capabilities is given in sec.~\ref{sec:overview}, followed by the
design choices, specifications, and fabrication details of the employed
diffractive optics in sec.~\ref{sec:optics}. In sec.~\ref{sec:controls},
we review the different control aspects necessary to collect data efficiently 
during an experiment. Finally, in sec.~\ref{sec:propagation}, we detail
the beam propagation calculator that we provide to users to design their
samples to be compatible with this setup.
In sec.~\ref{sec:analysis}, we detail the data analysis steps required to make the
best use of the collected data. First, we introduce basic statistical concepts
in sec.~\ref{sec:statistics}. In sec.~\ref{sec:imaging}, we discuss the imaged
beam on the DSSC detector, which leads to the flat-field correction
in sec.~\ref{sec:ff}. In sec.~\ref{sec:nl}, we describe the non-linear
correction and how it is calculated and applied to the data. In
sec.~\ref{sec:shotnoise}, we discuss how close these different corrections bring us
to the photon shot-noise limit. In sec~\ref{sec:offline}, we describe the offline analysis
procedure available to the users and in sec~\ref{sec:online}, we describe the tools
we provide for the analysis during the experiment as the data are being collected.
In sec.~\ref{sec:results}, we showcase some examples of experimental results,
starting in sec.~\ref{sec:trXAS} with the transient XAS in NiO and the impacts
of the different corrections on the data. Finally, in sec.~\ref{sec:limits},
we discuss the sensitivity limits of the current setup in terms of X-ray fluence
(sec.~\ref{sec:Xrayfluence}), repetition rate (sec.~\ref{sec:reprate}) and
typical sample systems that can and cannot be measured currently
with this setup (sec.~\ref{sec:sensitivity}).

\section{Setup\label{sec:setup}}

\subsection{Overview\label{sec:overview}}

\begin{figure}
\caption{Scheme of the beam-splitting off-axis zone plate (BOZ) setup.
  SASE FEL pulses are produced in the SASE3 undulators (UND). A variable
  line spacing plane grating (MONO) is placed 130~m downstream and disperses the
  SASE pulses on the exit slit (ES) placed 100~m further downstream. The
  monochromatic X-rays then propagate 30~m further and are split into three beams
  of equal intensity and focused down in front of the sample by the
  beam-splitting off-axis zone plate (BOZ).
  The sample (SAM) consists of three X-ray
  transparent membranes that are aligned on these three beams. The X-ray spot size
  on the membrane is typically 50$\times$50~$\mu$m$^2$. The middle membrane is used as a
  reference and is either a clear aperture or a bare membrane, while the other
  membranes contain the sample of interest, here a NiO thin film. The three beams further
  propagate and expand onto one sensor of the DSSC detector placed 5.4~m
  downstream. At the bottom of the figure, the resulting XAS spectra at the Ni
  L$_3$ edge are shown for the unexcited sample on the left, the sample excited
  by the optical laser (OL) on the right, as well as their measured difference
  in the middle. Above each of the three XAS plot, we identify the beam used as
  reference (I$_\mathrm{o}$) and the beam used as transmission (I$_\mathrm{t}$).
  During an energy scan, the DOOCS and Karabo control systems ensure that the gap size of
  the 120~m long undulator follows the monochromator photon energy. Similarly,
  the zone plate position is adjusted with the photon energy by typically
  a few millimeters to keep the zone-plate focus fixed in space.}
\label{fig:setup}
\includegraphics[width=\figuredoublewidth]{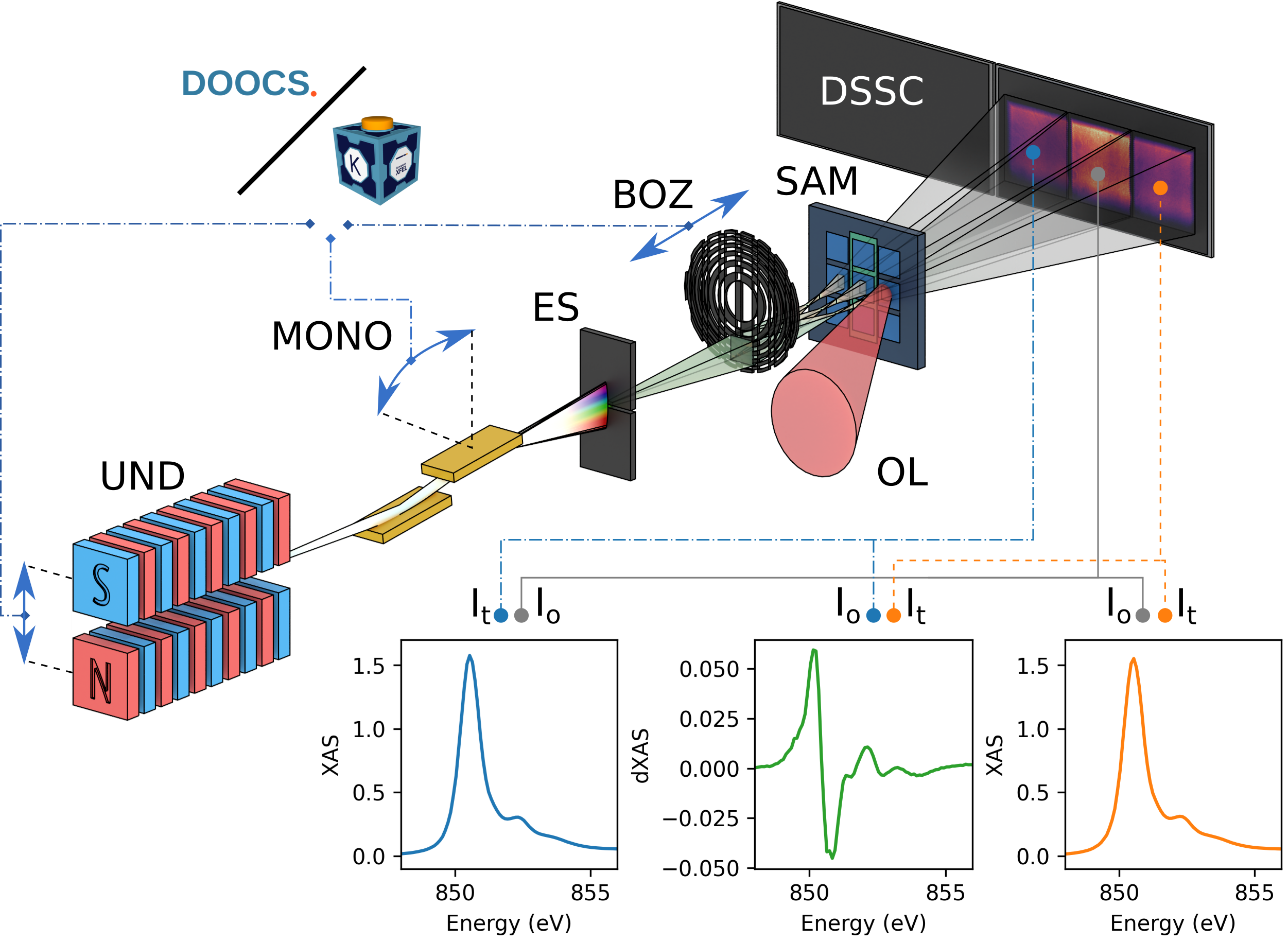}
\end{figure}

The setup implemented at the SCS instrument at the European XFEL is
schematically shown in Fig.~\ref{fig:setup}. The X-ray pulses are generated in
the SASE3 undulator system (UND). The X-ray photon energy is determined by the fixed
electron bunch acceleration energy and the variable undulator gap. The X-rays
are then monochromatized with the help of a variable line spacing grating (MONO)
combined with an exit slit (ES) \cite{Gerasimova2022a}. The monochromatic X-ray
pulses propagate through the beam splitting off-axis zone plate (BOZ)
optics. It consists of a transmission grating and a focusing
zone plate in a single element. The grating splits the initial beam into three beams of approximately
equal intensity. The zone plate focuses these beams shortly before the sample
(SAM), after which the beams then expand further downstream and are detected on a single
monolithic sensor of the DSSC detector, which is 3~cm high and 6.2~cm wide and
populated by 128 by 256 pixels \cite{Porro2021}. The high sensitivity of this
measurement scheme is assured by using a low-noise detector and illuminating
many pixels to achieve a high signal-to-noise ratio. The method's sensitivity is
then mainly limited by the number of photons detected, as we will show. The DSSC
can record up to 800 frames at 4.5 MHz during the FEL train, meaning an
effective repetition rate of 8~kHz.
The sample consists of an array of X-ray transparent
membranes, with each of the three X-ray beams passing through a separate membrane
window. The middle membrane window consists of the bare substrate, while the
right and left membranes each consist of the thin film under
investigation. To record a spectrum,  the X-ray photon energy can be scanned by varying
together the undulator gap, the monochromator energy, and the BOZ position
along the X-ray beam, with
the help of the Distributed Object Oriented Control System (DOOCS) \cite{Grygiel1996}
and Karabo control systems \cite{Hauf2019}, as shown in Fig.~\ref{fig:setup}
by dash-dotted lines connecting blue double headed arrows. Finally, for
stroboscopic or single-shot pump-probe experiments, an optical pump laser (OL) can be focused
onto one of the membranes to excite it \cite{Pergament2016}. The time delay between the optical pump
and X-ray probe pulse can be controlled by an optical delay line (not shown in
Fig.~\ref{fig:setup} for simplicity). By defining regions of interest (ROIs), the
transmitted intensity of  each of the three separated beams can be computed from
the DSSC
detector images. The XAS of the unexcited sample can be determined 
from the intensity of the beam going through the bare membrane (grating 0\ts{th} order) and
the beam passing through the unexcited sample (grating +1\ts{st} order), as shown
by the blue curve in the bottom left plot in Fig.~\ref{fig:setup}. Similarly, the
XAS of the excited sample can be determined simultaneously, as shown by the
orange curve in the bottom right plot in Fig.~\ref{fig:setup}. Finally, the pump-induced
XAS change can be determined, as well simultaneously, from the
intensity of the beam going through the unexcited sample and the beam going
through the excited sample, as shown by the green curve in the bottom center plot
in Fig.~\ref{fig:setup}. In the next sections, we will present
the design and fabrication of the BOZ optics,
the different control aspects necessary to collect a spectrum,
and finally tools available to users to aid in the design
of samples for this setup.


\subsection{Diffractive optics \label{sec:optics}}

Horizontally, the grating structure of the BOZ optics splits the beam into different
orders. The period of the grating structure is chosen to provide an
angular separation of the diffraction orders of 3.1~mrad, which is just
sufficient to prevent the beams from overlapping as they propagate 5.4~m up to
the DSSC detector placed at the end of the SCS experimental hutch. This gives, for example,
a grating structure period of 465~nm to operate around the Ni L$_{3,2}$ edges at
860~eV.

To detect as many photons as possible without saturating the detector, the
beams have to be as large as feasible. This means that the focal length
of the Fresnel zone plate component of the BOZ optics should be chosen as small
as possible. For this setup, we have chosen a focal length of 250~mm to ensure
that the sample can be placed just upstream of the zone-plate focus in its most
upstream position. Then, by using the 190~mm scanning range of the sample along the
beam propagation, we can control the spot size of the X-rays on the sample,
from tight focus at the zone plate focus to much larger beams in the most
downstream position. Similarly to the grating structure, the Fresnel zone plate
structure creates diffraction orders. Considering only the lowest diffraction
orders, we have the +1\ts{st} zone plate order, which is focused downstream at
the zone plate focus, the 0\ts{th} zone plate order, which is unfocused and simply
propagates through, and the -1\ts{st} zone plate order, which is diverging. With an
on-axis zone plate, all these different orders would spatially overlap on the
detector \cite{Schlotter2020}. Here we use an off-axis part of the zone plate, as
schematically shown in Fig.~\ref{fig:setup}, to vertically separate the different
orders on the detector. In our setup, this off-axis component, measured as the
distance between the BOZ optics center and the optical axis for the zone plate,
is chosen to be 0.55~mm. For the Ni L$_{3,2}$ edges at 860~eV, this gives an
off-axis Fresnel zone plate structure with an outermost zone width of 179~nm
for a zone plate aperture of 0.8$\times$0.8~mm$^2$.

The intensity ratio of the three focused beams can be controlled by two different
design parameters in the BOZ pattern as described: the pattern-inversion method
and the pattern-shift method \cite{Doering2020}. We applied the latter method with
a shift parameter of $s = 0.32$, as it gives higher overall efficiency.

The BOZ elements were made from single-crystal silicon
membranes \cite{Doering2020}. The 1~$\mu$m thick silicon membranes (Norcada Inc.,
Edmonton, Canada) were sputter-coated with a 10~nm chromium layer and then
spin-coated with a 70~nm polymethylmethacrylate (PMMA) resist. After the
electron-beam lithography (Vistec EBPG5000+, operated at 100~keV electron
energy) of the BOZ patterns and subsequent development, the resist patterns were
transferred into the Cr mask by reactive ion etching (RIE) in a Cl$_2$/O$_2$
plasma. After removal of the PMMA resist in acetone, the pattern was etched down
to about 700~nm deep into the silicon membranes by RIE in a
SF$_6$/C$_4$F$_8$/O$_2$ plasma. Finally, the Cr mask layer was removed to yield
pure silicon structures.

\subsection{Controls\label{sec:controls}}

To record a XAS spectrum, the monochromator~\cite{Gerasimova2022a} is scanned
continuously back and forth between two energy endpoints. The 120~m long undulator
system is controlled through the DOOCS control system \cite{karabekyan2012, karabekyan:icalepcs13-thppc095}.
A DOOCS middlelayer (ML) server provides an interface to specify the undulator photon energy and in turn controls
the gap size of each undulator. During a scan, the Karabo control system ensures the
undulator photon energy follows the monochromator photon energy through a feedback loop
which interfaces the DOOCS ML server through Karabo-DOOCS bridging software.
This combination allows recording spectra covering tens of eV.

However, to maintain full lasing of the undulators, the relative variation in the undulator deflection parameter
$\Delta K/K$, and respectively the relative change in the magnetic field strength
$\Delta B/B$, for the undulator system should not exceed the Pierce or FEL parameter
$\rho$ \cite{Pierce1950, Bonifacio1984, McNeil2010}. In the case of the SASE3 undulator system,
this relative change of the deflecting parameter $\Delta K/K$ should not exceed the value
of \num{1e-3}. The magnitude of the magnetic field depends on the undulator gap $g$ and the
undulator period $\lambda_u$ of the undulator is described by an exponential decay according to the expression:
\begin{equation}
    B\left(\frac{g}{\lambda_u}\right)[T] = a\exp{\left(b\frac{g}{\lambda_u} + c\left(\frac{g}{\lambda_u}\right)^2\right)},
    \label{eq:B}
\end{equation}
where, in the case of U68 undulator with period $\lambda_u$ = 68~mm, the parameters have the
following values a = 3.214, b = -4.623 and c = 0.925. By applying the partial differential
method with respect to changes of the gap $\Delta g$, the boundary condition for full
lasing can be determined:
\begin{equation}
    \frac{\Delta K}{K} = \frac{\Delta B}{B} = \Delta g\left(\frac{b}{\lambda_u} + 2c\frac{g}{\lambda_u^2}\right) \leq \rho,
\end{equation}
giving:
\begin{equation}
    \Delta g \leq \frac{\rho}{\frac{b}{\lambda_u} + 2c\frac{g}{\lambda_u^2}}.
    \label{eq:delta_g}
\end{equation}
Assuming that the working range of the U68 undulator gap is 10 to 25~mm,
the maximum deviation of the gap between undulators of one system should
not exceed $\Delta g$ = 15 to 17~$\mu$m to maintain the full lasing condition,
respectively.

If the undulator system is set to follow the monochromator with a scanning speed of 1~eV/s,
we need to determine the undulator gap scanning speed. From the undulator resonance equation for the first harmonic at
small observation angles:
\begin{equation}
    E[eV] = \frac{\num{2.48e-3}\gamma^2}{\lambda_u[mm]\left(1+\frac{K^2}{2}\right)},
    \label{eq:und}
\end{equation}
where $\gamma$ is the relativistic Lorentz factor for the electron, we can use the
the partial differential method with respect to the undulator gap change. The resulting gap
velocity values for 10~mm and 25~mm undulator gap are 10~$\mu$m/s and 2~$\mu$m/s respectively.
Measurements made on a system of four undulators showed that even without forced synchronization
of the axes of the undulators, the maximum deviation for a gap velocity of
0.856~mm/s corresponds to 40~$\mu$m \cite{karabekyan:icalepcs13-thppc095}.
It was also shown that this dependence is close to linear. By linear approximation for a gap velocity of 10~$\mu$m/s,
it could be concluded that the maximum deviation of the gap of undulators in one system will not exceed 0.5~$\mu$m.
This value confirms that by coupling the monochromator axis and the gap axes of the undulator system,
the full on-the-fly lasing condition for soft X-ray beamlines can be achieved, even with a few seconds delay in
communication between the undulator system and the monochromator.

One drawback of employing diffractive optics is that their properties are wavelength dependent.
For the BOZ, this means that both the focal distance and the grating diffraction angle
are proportional to the photon energy. During an extended energy scan, both the X-ray
spot size and the beam pointing on the sample can vary significantly. To compensate for these
effects, we use a three-axis linear piezo-motor stage to displace the BOZ along the X-ray beam
to a position calculated from the monochromator readback energy.
This ensures that the X-ray spot size and position on the sample remain constant during
the energy scan. The change in BOZ position along the X-ray beam $\Delta z$
due to a change in photon energy $\Delta E$ can be calculated
from the zone plate focal length $f$ at the design energy $E_0$ using:
\begin{equation}
    \Delta z = f\frac{\Delta E}{E_0}.
    \label{eq:BOZ-Delta_z}
\end{equation}
As an example, considering an energy scan spanning both Ni L$_{3,2}$
edges from 845~eV to 875~eV for a zone plate with a design energy of $E_0$ = 860~eV
with a focal length of $f$ = 230~mm, the change in BOZ position along the X-ray beam
$\Delta z$ is 8~mm.

\subsection{Beam propagation\label{sec:propagation}}

In contrast to the setup of~\cite{Schlotter2020}, where independent
manipulators were used to align individual sample and reference membranes in the
beam, the setup at the SCS instrument has only one sample manipulator.
Therefore, the sample has to be precisely designed to fit with the three beam
geometry from the start. To facilitate this, we publicly provide the \textbf{BOZcalc}
Python package to calculate the beam propagation and display projections at the
sample and detector plane~\cite{BOZcalc-rtd}.

\begin{figure}
\caption{a) Horizontal and b) vertical beam propagation from the intermediate
  source point to the detector. The position of the sample is shown as a dashed
  vertical line. The inset in each figure shows a zoomed-in region around the
  sample position. The red-filled areas represent the 707~eV photon beams from
  the fundamental harmonic of the undulators. The blue-filled areas represent the beams
  at 1414~eV photon energy originating from the second harmonic of the undulators. The
  vertical separation between fundamental and second harmonic beams near the
  zone-plate focus arising from the off-axis component of the diffractive optics
  can be used to block the unwanted radiation.}
\label{fig:geo_beams}
\includegraphics[width=\figurewidth]{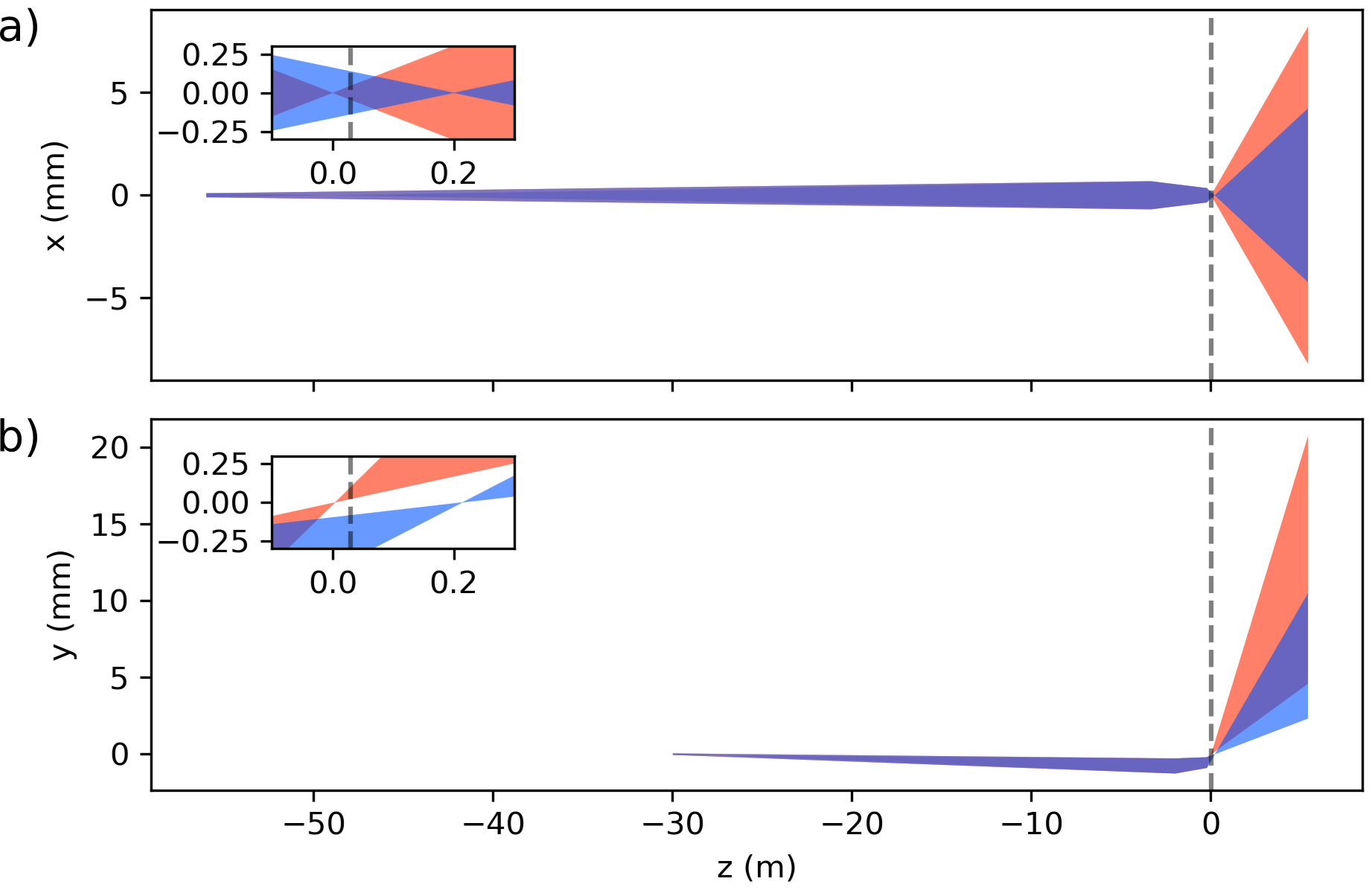}
\end{figure}

For example, in Fig.~\ref{fig:geo_beams}, the horizontal and vertical beam
profiles as a function of the distance from the interaction point, starting from
the horizontal and vertical intermediate source points
up to the DSSC detector, are displayed.
From the intermediate source point, the X-ray beam propagates down to the SCS
instrument and is slightly
focused by the Kirkpatrick-Baez (KB) mirrors~\cite{Mercurio2022} up to the BOZ. After that, the
beam is strongly focused and then expands until reaching the DSSC detector. The advantage of
using a zone plate with an off-axis component becomes evident in
Fig.~\ref{fig:geo_beams}~b), where the undulators' fundamental (red) and the second
harmonic (blue) are spatially separated. This allows using the sample as an
harmonic sorting aperture, blocking the unwanted undulator second harmonic
contribution that is not always suppressed by the beamline offset mirrors.

\begin{figure}
\caption{BOZ calculator showing the beams at the sample and detector planes as well
  as the control settings below. The focused 1\ts{st} zone-plate order beam footprint
  at the respective positions is shown as red for the funamental and blue for the second harmonic.
  The 0\ts{th} zone plate order is shown as gray. Individual membrane windows and the etching lines are shown
  in the sample plane figure as continuous and dashed lines, respectively. In the
  detector plane figure, the DSSC sensor is represented by the black rectangle
  and the DSSC filter mount is represented by the green shapes.
  Dimensions in both figures are in millimeters.}
\label{fig:bozcalc}
\includegraphics[width=\figurewidth]{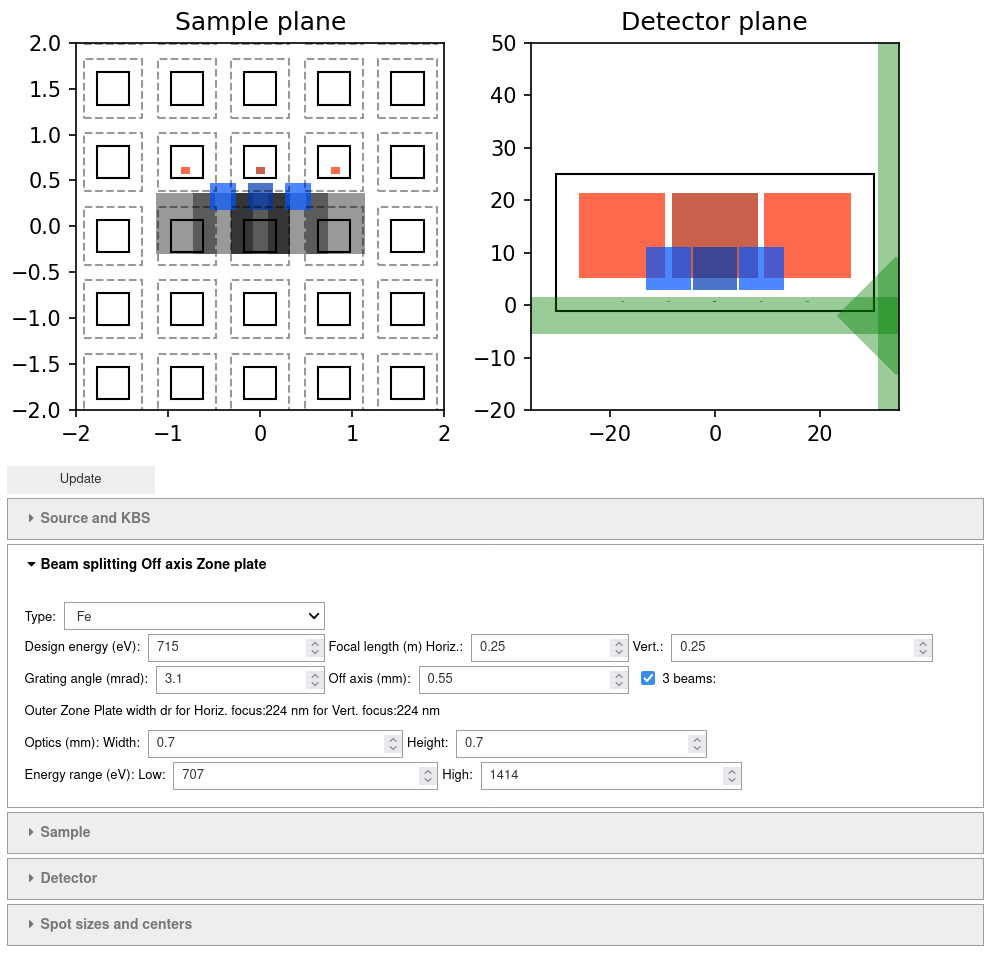}
\end{figure}

In Fig.~\ref{fig:bozcalc}, the results of a calculation of the
three beams positions and shapes, at the sample plane and detector plane, are shown in adjacent
figures side by side, as well as the control widgets at the bottom. In both
figures, the dimensions are in millimeters. All the calculations are performed
within a Jupyter notebook. In the sample plane figure, a
membrane array is displayed as a series of black rectangles. The etching facets
on the back of the substrate are represented as gray dashed lines. The red
squares indicate the expected beam size for the focused 1\ts{st} order beam of the
zone plate at the given sample position. The overlapping grey squares represent
the unfocused 0\ts{th} order of the zone plate. The blue squares are the expected
beam position and size of the second-harmonic beam from the undulator. In this
arrangement, the unwanted second-harmonic radiation would be blocked by the
sample frame, while the fundamental beam would propagate through the sample
membranes. On the detector plane, the black rectangle represents the DSSC module
used to record the beams' intensity, while the green shapes represent the DSSC
filter mount and are opaque to X-rays. The red squares show how the three beams
expand before reaching the DSSC to fill the sensor area after being focused by the zone plate
in front of the sample. It is worth noticing that the undulator second harmonics
would overlap with the other beams of interest, if they were not blocked by the
sample. It is worth mentioning that the grating 2\ts{nd} order, zone
plate 2\ts{nd} order of the second undulator harmonic will exactly overlap with
the grating 1\ts{st} order, zone plate 1\ts{st} order of the undulator fundamental and
therefore cannot be separated with an aperture. However, the
intensity of the 2\ts{nd} grating order is determined by the duty cycle of the grating
structure and is exactly zero for an even duty cycle.
In consequence, the pattern shift method shows the important advantage of suppressing
a potential contamination of the signal by the second harmonic beams appearing in
the 2\ts{nd} grating order \cite{Doering2020}.
The unfocused beams from the 0\ts{th} zone plate order are the barely visible
small spots seen on the filter mount near the bottom of the DSSC module. It is
also important to block these beams as they would otherwise saturate the
illuminated pixels. The accordion widgets below the figure show the
input controls of the calculator and can be interactively adjusted, for example,
to check the expected beam size at a given sample distance for a given zone
plate at a given photon energy.

Having presented the setup and the tools available to users to design samples
compatible with this setup, we will now detail the data processing and analysis
required.

\section{Data processing\label{sec:analysis}}

\subsection{Statistics\label{sec:statistics}}

The absorption of light propagating through a sample with thickness
$d$ is given by the Beer--Lambert law:
\begin{equation}
    I_\mathrm{t} = I_\mathrm{o} \mathrm{e}^{-\mu d},
    \label{eq:Beer-Lambert}
\end{equation}
where $I_\mathrm{o}$ is the incoming photon intensity, $I_\mathrm{t}$ is the transmitted
photon intensity after the sample, $\mu$ is the inverse of the absorption
length, and $d$ the sample thickness. The sample transmission is
$T = I_\mathrm{t}/I_\mathrm{o}$, and the X-ray absorption $A$ is:
\begin{equation}
    A = -\ln{\frac{I_\mathrm{t}}{I_\mathrm{o}}} = \mu d.
    \label{eq:XAS}
\end{equation}

The other important quantity to determine is the signal-to-noise ratio (SNR)
of the measurement, which is given by propagating uncertainties \cite{Meija2007, Schlotter2020}:
\begin{equation}
    \frac{|\mathrm{A}|}{\sigma_\mathrm{A}} = \frac{|\mathrm{T}|}{\sigma_\mathrm{T}} = \left. 1 \middle/ \sqrt{
    \left(\frac{\sigma_{I_\mathrm{t}}}{I_\mathrm{t}}\right)^2 + 
    \left(\frac{\sigma_{I_\mathrm{o}}}{I_\mathrm{o}}\right)^2 
    - 2 \frac{\sigma_{I_\mathrm{t}I_\mathrm{o}}}{I_\mathrm{t}I_\mathrm{o}}
    } \right.,
    \label{eq:SNRratio}
\end{equation}
where $|\;|$ denotes the absolute value, $\sigma$ is the standard deviation, and $\sigma_{I_\mathrm{t}I_\mathrm{o}}$ denotes the
covariance.
In the photon shot-noise limit, we have $\sigma_{I_\mathrm{o,t}} = \sqrt{N_\mathrm{o,t}}$, where $N_\mathrm{o,t}$
is the number of photons. For a given $I_\mathrm{o}$ intensity, the photon shot-noise uncertainties are uncorrelated such that
$\sigma_{I_\mathrm{t}I_\mathrm{o}}$ is 0. Simplifying eq.~\eqref{eq:SNRratio} in the photon shot-noise
limit gives:
\begin{equation}
\mathrm{SNR} = \sqrt{\frac{\mathrm{T}\mathrm{N}}{1 + \mathrm{T}}},
\label{eq:shotnoiselimit}
\end{equation}
where $N$ is the number of photons in the $I_\mathrm{o}$ beam and $T$ is the
average transmission. For a weakly absorbing sample with a transmission close to 1,
the SNR limit is thus $\sqrt{N/2}$, while for a strongly absorbing sample, i.e. with a transmission
of 0.1, the SNR limit will be dominated by the noise in the transmitted beam with less photons, giving a
reduced SNR limit of $\sqrt{N/10}$ in this example.

When data are recorded, several shots are taken at the same photon energies (or time delay)
and have to be averaged together. The first approach is to compute the transmission $T_i$ for
each of the $M$ shots and average them together, such that $T = 1/M\sum_{i=1}^{M} T_i$. The uncertainty
is simply given by the standard deviation $\sigma_T$ of the $T_i$ and the single-shot SNR is:
\begin{equation}
    \mathrm{SNR} = \frac{T}{\sigma_T}.
    \label{eq:SNR}
\end{equation}

However, shots with weak intensity, due to the high fluctuations introduced by monochromatizing the SASE pulses,
tend to be noisier and dominate the uncertainty in the measurement. A better approach is to first
compute the summed incoming
$\sum_i {I_\mathrm{o}}_i$ and transmitted $\sum_i {I_\mathrm{t}}_i$ intensities before calculating the
averaged transmission $T_\mathrm{w}$:
\begin{equation}
    T_\mathrm{w} = \frac{\sum_i {I_\mathrm{t}}_i}{\sum_i {I_\mathrm{o}}_i}.
    \label{eq:sum_intensity}
\end{equation}
Here, we are simply summing up the photons detected over many shots together. This is identical to
computing a weighted average of $T_i$ with ${I_\mathrm{o}}_i$ as weight, i.e. intense shots should
contribute more and weak shots should contribute less to the mean. The uncertainty in the measurement is now given
by the weighted standard deviation:
\begin{equation}
    \sigma_\mathrm{w} = \sqrt{\frac{\sum_i {I_\mathrm{o}}_i {(T_i - T_\mathrm{w})}^2}{V_1 - V_2/V_1}},
    \label{eq:wsigma}
\end{equation}
with $V_1 = \sum_i {I_\mathrm{o}}_i$ and $V_2 = \sum_i {{I_\mathrm{o}}_i}^2$. The single-shot weighted SNR is then:
\begin{equation}
    \mathrm{SNR}_\mathrm{w} = \frac{T_\mathrm{w}}{\sigma_\mathrm{w}}.
    \label{eq:SNRw}
\end{equation}

\subsection{Imaging\label{sec:imaging}}

\begin{figure}
\caption{a) Dark-corrected single-pulse image of the DSSC sensor showing the three
  imaged beams (-1\ts{st}, 0\ts{th} and +1\ts{st} order).
  b) Dark-corrected average image of a single DSSC sensor showing
  the three imaged beams based on a 5 minute long data acquisition recording 3000
  FEL trains with 15 X-ray pulses per train. c) Image showing each beam
  normalized by the 0\ts{th} order beam. d) Fitted flat-field correction. e) Dark-corrected
  and flat-field corrected average image in a single DSSC sensor. The
  three imaged beams now appear as identical copies of the initial beam.}
\label{fig:ff}
\includegraphics[width=\figurewidth]{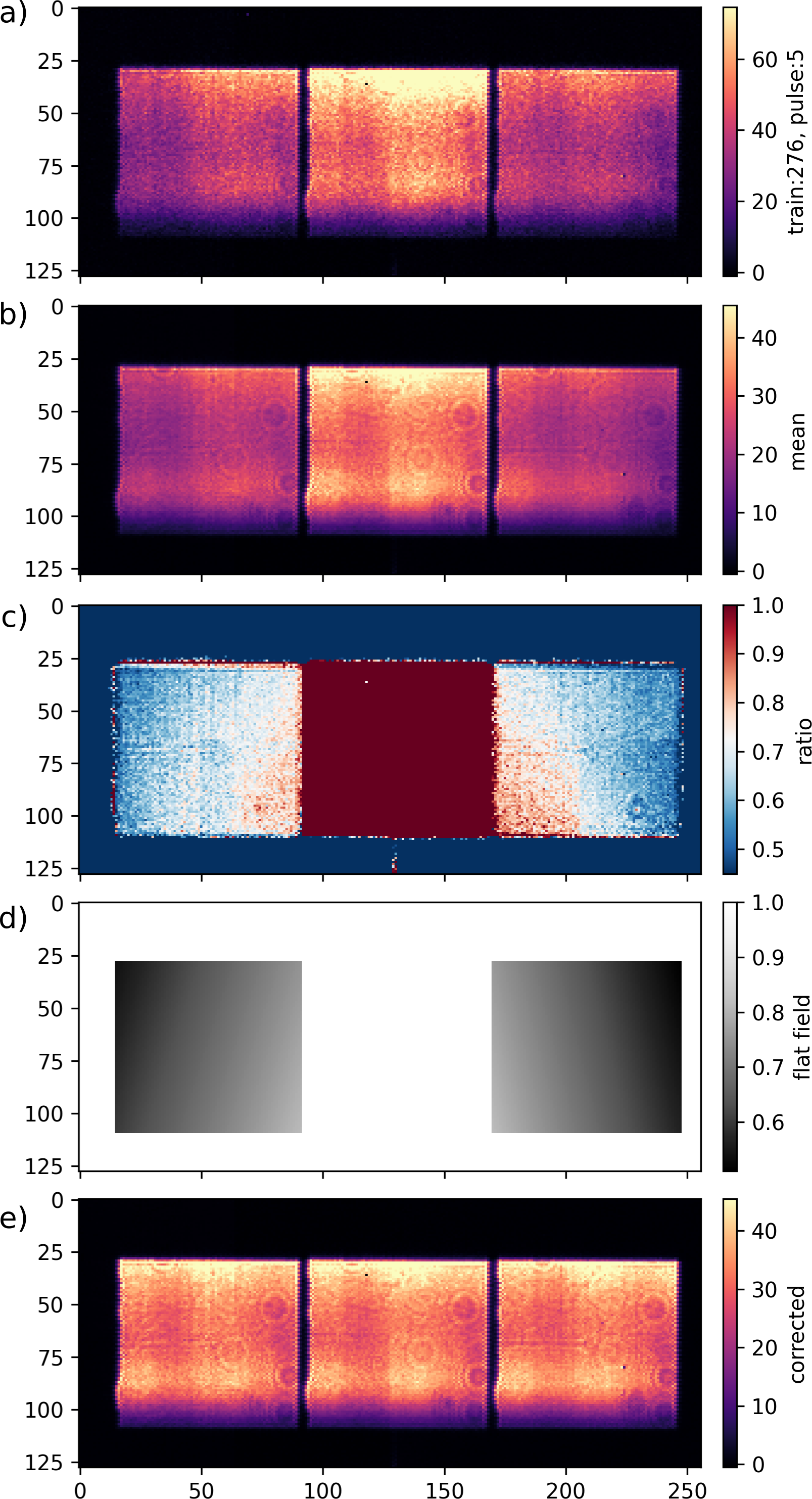}
\end{figure}

To characterize the sensitivity of the setup, we collect a set of data without samples in the beam,
in which case the transmission is known to be exactly $T = 1$. In Fig.~\ref{fig:ff}~a), a
single SASE pulse is imaged on the DSSC sensor, showing the three characteristic beams from the
diffractive optics. To simplify the discussion in the rest of the manuscript,
we identify these three beams by their -1\ts{st}, 0\ts{th} and +1\ts{st} grating order alone,
omitting to mention their +1\ts{st} zone plate order part.
The shape of each beam is roughly a square and is determined by the
aperture size of the zone-plate optics. A four-slits system directly upstream
of the diffractive optics is used to control the incoming beam size and to ensure separation
of the diffracted beams on the DSSC detector. The average dark-corrected image over a 5 minutes long dataset
consisting of 3000 FEL trains with 15 X-ray pulses per train is shown
in Fig.~\ref{fig:ff}~b). The first point to note is that both the -1\ts{st} and +1\ts{st} grating orders are
somewhat weaker than the 0\ts{th} grating order in the middle. This is due to the design of the diffractive optics and the
resulting diffraction efficiency in each order, which can be tuned \cite{Doering2020}.
The second point to note is that the beam intensity of the right beam decreases on its right side, and similarly,
the left beam intensity decreases on its left side. We suspect this effect to be an intrinsic property of the
diffractive optics and not a fabrication issue, as it occurs for all the optics we tested. 
The variation of intensity across different beams makes the measured transmission spatially dependent
and deviating from 1.0. In the next section, we detail how this effect can be corrected.

\subsection{Flat-field correction\label{sec:ff}}

The ratio between each beam and the 0\ts{th} grating order beam are shown
in Fig.~\ref{fig:ff}~c). One notices the presence of gradients
in the -1\ts{st}/0\ts{th} and the +1\ts{st}/0\ts{th} ratios, with values ranging from 0.55 up to 0.85.
The orientation of the
gradients is along the diagonal direction with respect to the square shape of the beams and they appear
to be mirror images of each other. To correct for the gradients, we fit a plane defined
by ${ax + by + cz + d = 0}$, where $x$ and $y$ are the horizontal and vertical
positions of a pixel in the sensor, respectively, and $z$ is the value of the
ratio computed for the corresponding pixel. We can impose a horizontal mirror symmetry
such that both gradients are fitted with only four fitting parameters, $a$, $b$, $c$, and $d$ in total.
The result of the fitting is shown in Fig.~\ref{fig:ff}~d), where a value of 1.0 is set everywhere, except
where the -1\ts{st} and +1\ts{st} beams are located, in which case the fitted plane and its mirror are evaluated.
Dividing the data shown in Fig.~\ref{fig:ff}~b) by the normalization shown in Fig.~\ref{fig:ff}~d)
results in the data shown in Fig.~\ref{fig:ff}~e). The imaged beams now appear as three identical copies of the same
initial beam image, as we would expect from the property of a grating. We call this normalization step the
flat-field correction.

\begin{figure}
\caption{Two-dimensional histograms of the three ratios (-1\ts{st}/0\ts{th} in the top row,
1\ts{st}/0\ts{th} in the middle row, and -1\ts{st}/1\ts{st} grating order in the bottom row) as a function of the number of
photons in the 0\ts{th} order for dark-corrected data in the left ``raw'' column, for the dark-corrected
and flat-field corrected data in the middle ``flat-field'' column and finally the dark-corrected,
flat-field corrected and non-linear corrected in the right ``non-linear'' column. Data that
contains saturated pixels are shown as red, as opposed to blue where no saturation occurs. 
In each plot, the SNR and weighted SNR (SNR$_\mathrm{w}$) of the non-saturated events are indicated.}
\label{fig:corr}
\includegraphics[width=\figurewidth]{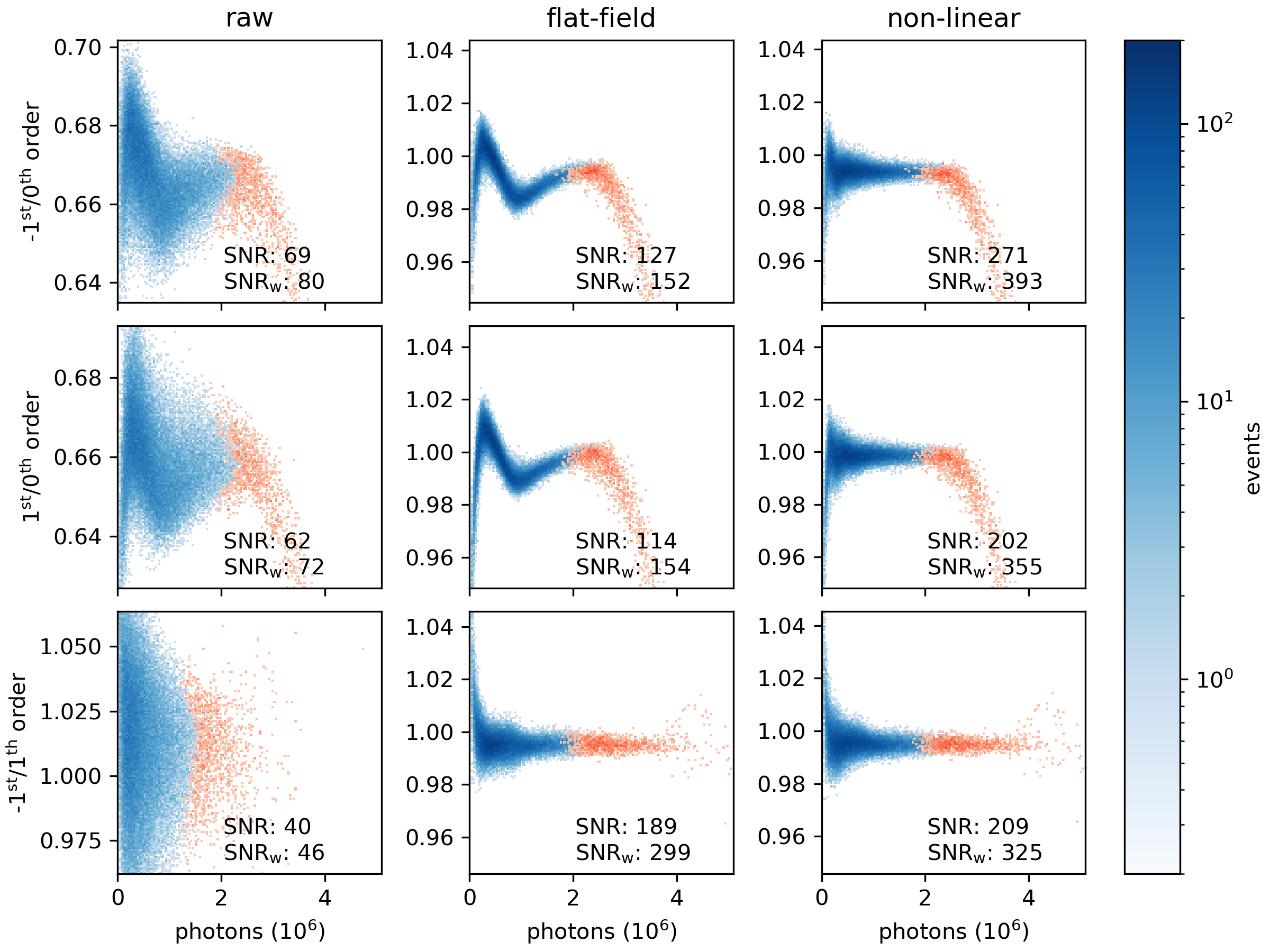}
\end{figure}

To better understand the statistical distribution of the data, we extract the intensity of the three beams
by defining three regions of interest (ROIs)
around each beam and summing up the measured pixel values within them after dark correction. We then
compute the three possible ratios -1\ts{st}/0\ts{th}, +1\ts{st}/0\ts{th}, and -1\ts{st}/+1\ts{st} grating orders as a
function of the intensity in the 0\ts{th} grating order for every pulse in every train in the dataset. The
resulting two-dimensional histograms are shown in Fig.~\ref{fig:corr}. The column on the
left in Fig.~\ref{fig:corr} labeled ``raw'' shows data that are only dark-corrected. We can see that
they are quite dispersed. In these plots, the data in red show the pulses where at least one pixel is
saturated and are thus not reliable, while the data
in blue show no saturation of the DSSC detector. On each plot, the SNR given by eq.~\eqref{eq:SNR} and
weighted SNR (SNR$_\mathrm{w}$) given by eq.~\eqref{eq:SNRw} are shown for the non-saturated data.
After applying the flat-field correction, which compensates for some diffraction-efficiency variation
in the diffractive optics, we obtain the data shown in the middle column in Fig.~\ref{fig:corr},
labeled ``flat-field''. We can see that the improvement is significant and that these first corrected data
are much less dispersed. We also see a large increase in SNR$_\mathrm{w}$ by a
factor of 6 for the -1\ts{st}/+1\ts{st} ratio, from 46 to 299. For the two other ratios, the improvement is more
moderate and only a factor of 2. In addition, the measured transmission now clearly appears non-linear
as a function of the intensity in the 0\ts{th} order, a feature that was in part hidden in the noise before.

In practice, the zone plate can be slightly rotated with respect
to the DSSC sensor. This can be taken into account in the fitting procedure of the flat-field correction
by lifting the horizontal symmetry requirement at the cost of doubling the number of fitting parameters from four to eight.
In addition, a ratio is only properly defined where we have enough intensity, and creates outliers otherwise.
The process of finding ROIs encompassing bright enough pixels and excluding outliers can be time consuming. To
solve that problem, a more reliable approach was derived. The idea is based on the change displayed by flat-field corrected
data as compared to the raw data and shown in Fig.~\ref{fig:corr}. If we were to bin the data in $k = 40$ small intervals
of intensity in the 0\ts{th} order, then, within each of these intervals, the spread in the data for each of the three ratios is reduced
by the flat-field correction. This spread is naturally measured with the standard deviation over each $k$ interval, $\sigma_k$.
To fit the flat-field correction, we introduce the following criterion $J_\mathrm{ff}$:
\begin{equation}
\begin{split}
    J_\mathrm{ff} = 10^3 \left[
        \alpha_\mathrm{ff} \sum_k \left(
            \sigma_k\left(\frac{-1^\mathrm{st}}{0^\mathrm{th}}\right) +
            \sigma_k\left(\frac{1^\mathrm{st}}{0^\mathrm{th}}\right) +
            \sigma_k\left(\frac{-1^\mathrm{st}}{1^\mathrm{th}}\right) \right) \right. \\
        + \left(1 - \alpha_\mathrm{ff}\right) \left(
            \left(1-<\frac{-1^\mathrm{st}}{0^\mathrm{th}}>\right)^2 +
            \left(1-<\frac{1^\mathrm{st}}{0^\mathrm{th}}>\right)^2 + \right. \\
            \left. \left(1-<\frac{-1^\mathrm{st}}{1^\mathrm{th}}>\right)^2 \right)
            \left. \vphantom{\sum_k} \right]
    \label{eq:Jff}
\end{split}
\end{equation}
where the term in $\alpha_\mathrm{ff}$ is the sum over all $k$ intervals of the standard deviation for all three ratios.
The term in $(1-\alpha_\mathrm{ff})$ is a regularization term to keep the mean $<>$ of each of the three ratios around unity.
Minimizing $J_\mathrm{ff}$ as a function of the eight fitting parameters (four for each plane without mirror symmetry)
proved to be more reliable
and gave better estimates of the flat-field correction compared to fitting a plane to the ratio
of the mean images of each beam. In practice, the fitting procedure converges within a few iteration
and a regularization parameter $\alpha_\mathrm{ff}$ of about 0.1 works well and is not critical.
By choosing the $k$ intervals to be small enough, we can be sure that the non-linear trend, visible in
Fig.~\ref{fig:corr} for the flat-field corrected data, is not contributing to $J_\mathrm{ff}$. While
the improvement brought in by the flat-field correction in the data is very significant, it is clear
that the non-linearity observed in Fig.~\ref{fig:corr} needs to be addressed, if we want to make the best
use of the available data.

\subsection{Non-linear correction\label{sec:nl}}

\begin{figure}
\caption{
Fitted non-linear deviation from the ideal detector response
$\mathrm{F}_{\mathrm{nl}}(x) - x$
as a function of the DSSC pixel values. Inset, the evolution of the weighted SNR (blue squares)
and the correction cost (orange circles) as a function of the fitting iteration number.
}
\label{fig:fit}
\includegraphics[width=\figurewidth]{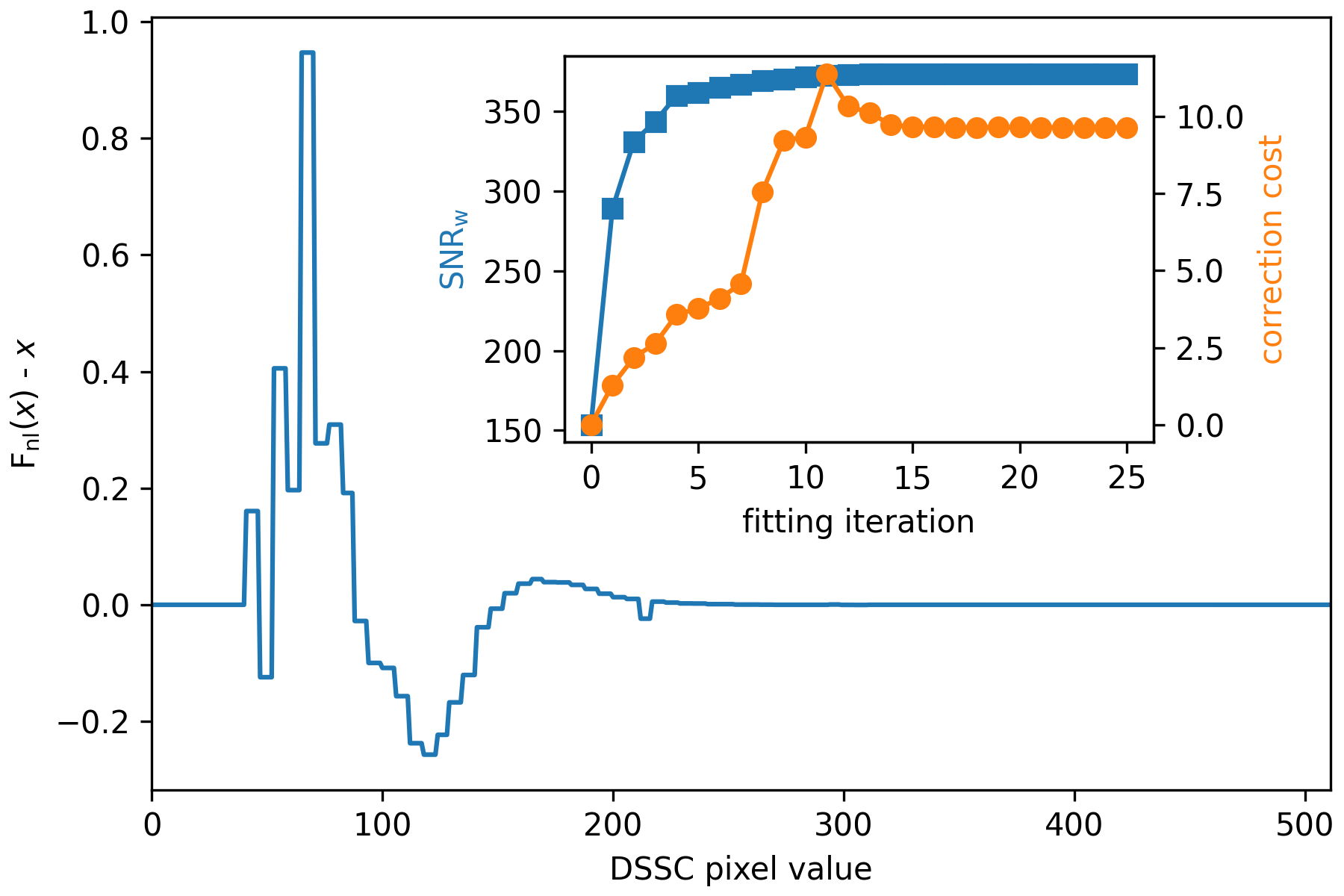}
\end{figure}

The mini-silicon drift detector (miniSDD) camera is a linear system, with an
analog chain linearity error better than 0.25\% \cite{Grande2019} and an
analog-to-digital converter (ADC) with integral non-linearity (INL) and
differential non linearity (DNL) better than 0.5 least significant bit (LSB)
and 0.32 LSB on the full range, respectively \cite{Hansen2013}. Nevertheless,
the remaining non-linearity that we observe needs to be addressed to further
improve the data quality.
Here we assume as a first approximation that the
DSSC single-pixel response is non-linear as a function of the incoming photon intensity. Moreover, we
assume that this non-linearity can be corrected by a \textit{pixel-independent} non-linear correction
function $F_\mathrm{nl}$ that only deviates slightly from the ideal linear detector response. We
model $F_\mathrm{nl}(x)$ over the integer range from 0 to 511,
representing the 9 bits of the DSSC pixel output values, as a piecewise constant function composed of $S$
segments. It starts from a user-defined low level $L$ and goes up to a high level $H$. In practice, we have
$S = 80$, $L = 40$, which is below the dark pedestal, and $H = 511$. To apply this correction function to
the collected raw data output of the DSSC, we proceed with the following algorithm:
\begin{enumerate}
    \item Replace integer value $x$ with float value $F_\mathrm{nl}(x)$ in both dark run data and run data.
    \item Subtract from the run data the pulse resolved mean dark value.
    \item Divide run data by flat-field normalization.
    \item Sum run data pixel values over each ROI.
\end{enumerate}

We then compute the weighted variance $\sigma^2_w$ according to eq.~\eqref{eq:wsigma}
for each of the three intensity ratios
measured in each ROI. The goal is then to fit the
\textit{pixel-independent} non-linear correction function $F_\mathrm{nl}$ in order to maximize the SNR$_\mathrm{w}$
of the -1\ts{st}/0\ts{th} and the +1\ts{st}/0\ts{th} ratio. For this, we calculate the following
criterion $J_\mathrm{nl}$:
\begin{equation}
    J_\mathrm{nl} = (1 - \alpha_\mathrm{nl})\frac{10^8}{2}(
        \sigma^2_w(\frac{-1^\mathrm{st}}{0^\mathrm{th}}) + \sigma^2_w(\frac{+1^\mathrm{st}}{0^\mathrm{th}}))
    + \alpha_\mathrm{nl} \sum_x (F_\mathrm{nl}(x) - x)^2
    \label{eq:Jnl}
\end{equation}
with $\alpha_\mathrm{nl}$ being a user defined parameter between 0 and 1 controlling the strength of the regularization
term. This term prevents the fitting from diverging to an unrealistic non-linear correction function by keeping
the correction cost, i.e. the deviation from the ideal detector response, as small as possible. In practice,
we use $\alpha_\mathrm{nl}$ = 0.5 as default value. We then minimize $J_\mathrm{nl}$ as a function of
the $S$ piecewise constant values modeling $F_\mathrm{nl}$. This computation typically takes
2 to 8 hours on a single node on the Maxwell computational resources operated at DESY and accessible
to the users of the European XFEL. In Fig.~\ref{fig:fit}, the fitted non-linear correction deviation
$F_\mathrm{nl}(x) - x$ is shown and is indeed small. Here, a maximum deviation of less than 1 for an input
value of about 80 is seen. As the DSSC was operated at a frame rate of 4.5~MHz, the actual input value recorded
in this dataset does not extend beyond 280 \cite{Porro2021}. This explains why the deviation is zero in the range of 280 to
511. The inset in Fig.~\ref{fig:fit} shows the evolution of
each component of the
minimization criterion $J_\mathrm{nl}$ as a function of the fitting iteration number, with SNR$_\mathrm{w}$ being
$1/\sqrt{10^{-8}J_\mathrm{nl}(\alpha_\mathrm{nl}=0)}$ and the correction cost being
$J_\mathrm{nl}(\alpha_\mathrm{nl}=1)$. One can see that within a few
iteration, the SNR$_\mathrm{w}$ increases significantly for a very moderate increase in the correction cost. With
further iteration, the SNR$_\mathrm{w}$ increases slightly to a plateau, at the cost of a much larger increase in the
correction cost. At this stage, the only gain in minimizing $J_\mathrm{nl}$ is by reducing the
correction cost, as shown by the small reduction around iteration 15 while the SNR$_\mathrm{w}$ remains constant.
Overall, within 25 iterations, the fitting has converged. In Fig.~\ref{fig:corr}, the data corrected for
dark, flat-field, and non-linear response are shown in the right column labeled ``non-linear''. It is
evident from these plots that the data are now much more linear, with the exception of the saturated data
in red, which are discarded from further analysis in any case. We see an increase by a factor of
2.5 in SNR$_\mathrm{w}$ with the addition of the non-linear correction for the -1\ts{st}/0\ts{th} and the
+1\ts{st}/0\ts{th} ratio. For the -1\ts{st}/+1\ts{st} ratio, the gain is much smaller. This is easy to
understand, considering that since the +1\ts{st} and -1\ts{st} are very similar in intensity, their
ratio is largely independent of detector non-linearity. This, in turn, motivates the omission of the -1\ts{st}/+1\ts{st} ratio
in eq.~(\ref{eq:Jnl}). Interestingly, we note that discarding the saturated pulses when computing
the -1\ts{st}/+1\ts{st} ratio might not always be the best strategy as these data do not apparently deviate
significantly from the non-saturated data. This could be due to the fact that the increase in beam
intensity is well determined by the many non-saturated pixels and not dominated by a few saturated pixels
in each beam, in contrast to beams of dissimilar intensity, where saturation occurs only in one of the beams.
In summary, combining a flat-field and a non-linear correction that can be efficiently calculated,
we significantly improved the collected
data. In the next section, we will discuss how close the corrected data are to the photon shot-noise limit.

\begin{figure}
\caption{SNR (Inverse of the standard deviation) of the data binned as a function of the
intensity in the 0\ts{th} order for a) the -1\ts{st}/0\ts{th} order, b) the +1\ts{st}/0\ts{th} order
and, c) the -1\ts{st}/+1\ts{th} order. Data that are only dark-corrected are shown as dotted blue
lines. Data that are also flat-field corrected are shown as dotted-dashed orange lines. Data that are also
corrected for non-linearity are shown as continuous green lines. The photon shot-noise limit is shown as dashed red lines.}
\label{fig:SNR}
\includegraphics[width=\figurewidth]{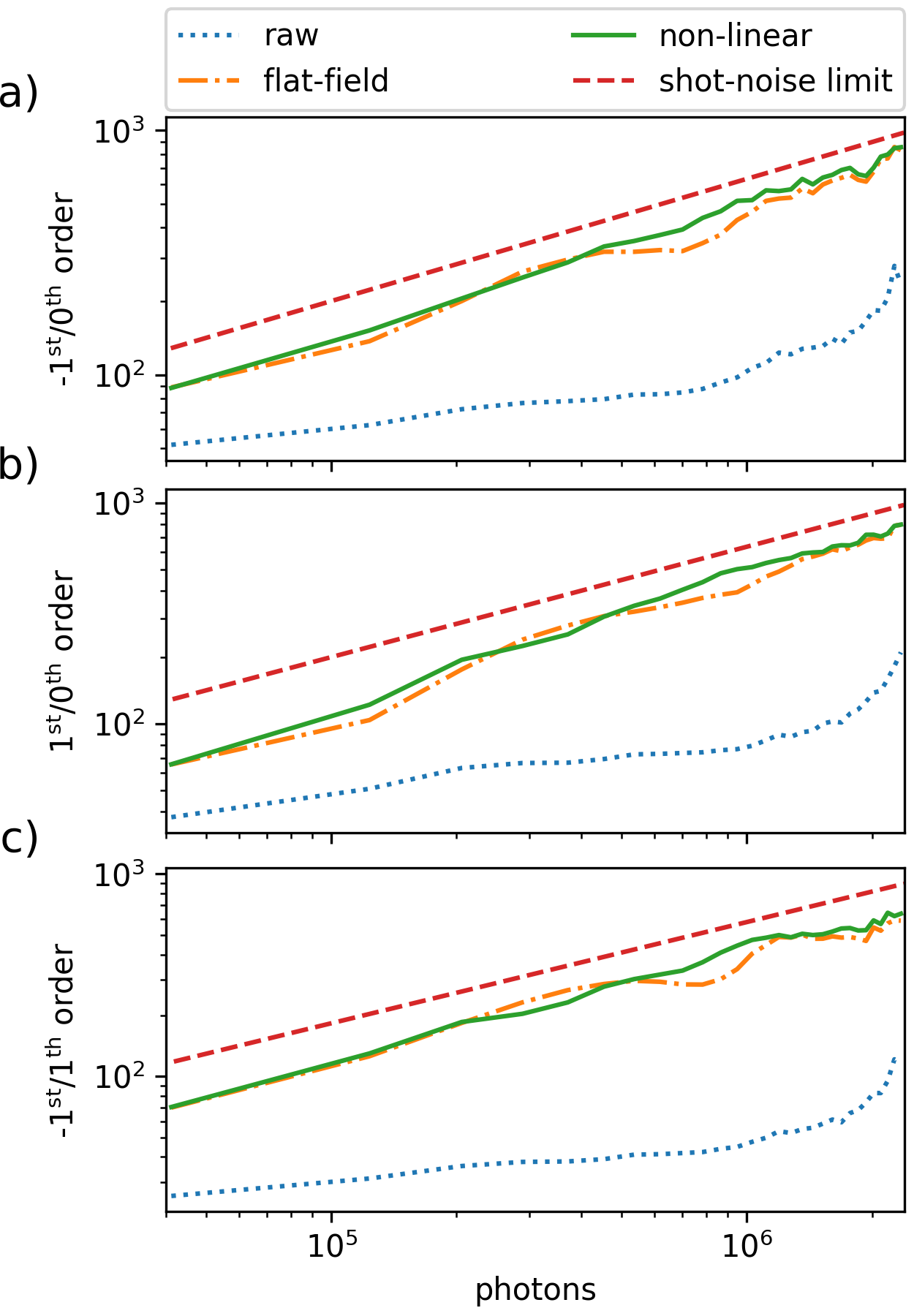}
\end{figure}

\subsection{Photon shot-noise limit\label{sec:shotnoise}}

To address the question of how close the corrected data are to the photon shot-noise limit, we plot in
Fig.~\ref{fig:SNR} the inverse of the standard deviation of the data binned as a function of the
intensity in the 0\ts{th} order. The data, which are only dark-corrected, are shown as blue
dotted lines. Data that are additionally flat-field corrected are shown as orange dash--dotted lines.
Data that are also non-linear corrected are shown as continuous green lines.
The photon shot-noise limit given by eq.~\eqref{eq:shotnoiselimit} is shown as red dashed lines in Fig.~\ref{fig:SNR}.
With the flat-field correction, data are already approaching the photon shot-noise
limit closely. The effect of correcting for non-linearity is not visible in this plot as we only plot the inverse of
the standard deviation of the binned data and not the systematic deviation from 1. From Fig.~\ref{fig:SNR}, we conclude that
we are making efficient use of every photon detected by the
DSSC detector, using the detailed data correction steps.

Before looking at actual time-resolved transient XAS measurements on the sample and confirm that the data treatment
gives sensible results, we discuss the different corrections we apply to the data and their origin.
At the moment, we lack a predictive model for the position-dependent diffraction efficiency of the zone plate,
which we correct with the flat-field correction. Nevertheless, it seems that the plane approximation to that
unknown dependence is sufficient.
For the non-linear correction, the miniSDD DSSC
pixel response is linear within
the expected margins \cite{Grande2019, Hansen2013, Porro2021}, but as we have shown, the data quality
can be further improved by correcting the remaining non-linear behaviour
with a \textit{pixel-independent} non-linear correction function.
Fortunately, the data processing that we detailed allows us to completely mitigate these effects and to reach the desired regime,
where the sensitivity of the setup is only limited by the number of detected photons.

\subsection{Offline analysis\label{sec:offline}}

The analysis procedure, which we detailed in the previous section, is made publicly available to all users as
routines in the SCS toolbox Python package~\cite{SCSToolbox-git}, with example Jupyter notebooks readily
available in the online
documentation~\cite{SCSToolbox-rtd}. The workflow is quite simple and detailed here. First, we use a dedicated notebook
to calculate the flat-field and non-linear correction on a set of data recorded without a sample.
This computation takes several hours for the non-linearity correction but only few tens of minutes for the flat-field
correction. The result is saved in a small JSON file that can be used later on to process data, both for
the offline analysis and for the online analysis. To speed up the data analysis during the beamtime, an intermediate
JSON correction file is saved as soon as the flat-field correction is finished. Second, the processing of data recorded
with a sample is split
into two parts, each having its dedicated notebook. The first part consists of processing the DSSC data,
applying all the detailed corrections and computing the intensity in each beam for each pulse in each train, and saving
these in an intermediate small data file in the proposal folder on the Maxwell
computational resources. The second part is to load one or several
of these small data files and to compute the XAS spectra or time delay traces with a binning procedure. This part can be easily modified
and adapted by the users to their needs during and after the beamtime.

Offline analysis programs and notebooks make use of European XFEL's Extra-data package \cite{Fangohr-2018} which provide convenient access
to the data files written at EuXFEL. The Python-based Extra-data framework~\cite{extra-data} makes data available through common data
science tools and objects such as numpy's arrays~\cite{numpy-2020}, xarray~\cite{xarray-2017} and dask array~\cite{dask}. In particular, it
is thanks to the multiprocessing capability offered by dask array that the computation time of the non-linearity correction could be reduced
from days to just few hours.

Jupyter notebooks are used by European XFEL users and staff to explore and analyse experiment data~\cite{fangohr-notebooks-data-exploration}.
The JupyterHub installation of the Maxwell cluster provides remote execution of Jupyter notebooks using the Maxwell resources and thus provides
an alternative to remote X, FastX or other remote access technologies. This is of particular value as the data sets recorded at the EuXFEL can be
so large, up to a Petabytes for a 5 days beamtime recording the full DSSC detector with 800 frames per train, that they typically stay at the facility and
need to be analysed remotely after the beam time. Here, for experiments employing a single DSSC module and recording few tens of pulses per train, the amount of data
generated is more moderate, in the order of few tens of Terabytes per beamtime. The use of Jupyter notebook can also help to make data analysis and publications more
reproducible~\cite{Beg-notebooks-reproducibility-2021}.

\subsection{Online analysis\label{sec:online}}

FEL beamtimes are both expensive and limited. It is therefore
crucial for users to be
able to
analyze the data in real-time in order to steer the experiment and maximize the
scientific output.
Karabo is designed to support concurrent initial analysis during data acquisition \cite{Hauf2019, Fangohr2018}.
It is a distributed software that consists of small pluggable components, so-called devices, that represent various components: a detector, a piece of equipment such as a sensor, or a control and analysis procedure such as a scanning routine. Karabo also includes a graphical user interface that allows feedback on the control system. 

There are different possibilities to achieve real-time data analysis at the
European XFEL, that is developing a related Karabo device that describes the analysis \cite{Flucke2020} or connecting an external application via a bridge \cite{Fangohr2018}. The DSSC detector produces up to 800 images per train
with a data rate of 1~GB/s for a single module. To ensure low-latency data processing, we have used EXtra-metro \cite{Metro-rtd}. It is a framework
developed in-house with intrinsic parallelization, which enables fast and
reliable online preview of various analysis routines.
These routines are generated by interpreting a Python script, where the analysis procedures are described. 

\begin{figure}
\centering
\includegraphics[width=0.8\textwidth]{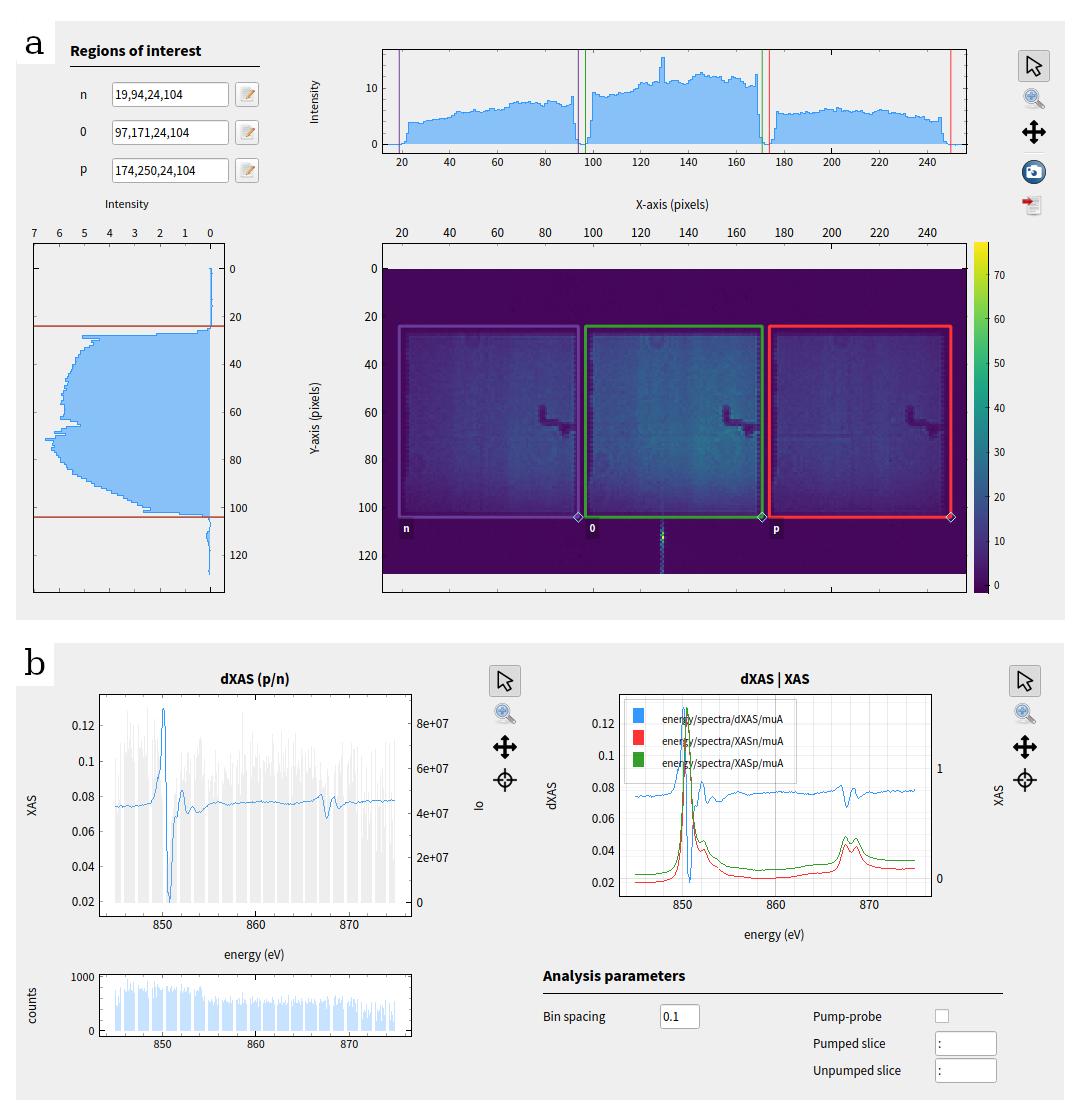}
\caption{a) Dark-corrected single-train image of the DSSC sensor in the Karabo GUI. The overlayed regions of interest define the intensity of the imaged beams
and can be modified using the EXtra-metro parameter fields or by user interaction
on the GUI. The projections along $x$- and $y$-axes are also plotted as guides for optimal zone plate alignment. b) XAS spectra of the three ratios after an energy scan displayed in the Karabo GUI. The panel also contains the analysis parameter fields that can be changed during runtime.}
\label{fig:online-plots}
\end{figure}

During the experiment, the detector images are recorded while scanning either the X-ray photon energy or the pump laser time delay parameters.
These data are simultaneously collected by EXtra-metro directly from the control system using
the data pipelines for large detector data and from the central messaging broker for the control data. The analysis routines defined in the
SCS toolbox package (sec.~\ref{sec:offline}) are then applied to the received data,
using the pre-calculated non-linear and flat-field corrections, in a train-by-train manner.

The processed data are then displayed in the Karabo graphical user interface. Custom widgets are developed for further data visualization and interaction using the GUI-extensions \cite{Flucke2020}. Fig.~\ref{fig:online-plots}~a) shows a
dark-corrected detector module with three overlayed regions of interest that define the intensities of the pumped, unpumped, and reference signals. These regions of interest, along with other analysis parameters such as bin spacing and pulse selection, can also be modified during runtime.
In addition to showing the DSSC image and ROIs, the vertical and horizontal projections of the intensity are displayed and are used during the initial zone plate alignment to find
the center of the X-ray beam. Finally, the resulting X-ray absorption spectra of the signals are shown in Fig.~\ref{fig:online-plots}~b) and are further discussed in the next section.

\section{Results\label{sec:results}}

\subsection{Transient XAS\label{sec:trXAS}}

It is essential to verify the effect of the different levels of data correction on actual
time-resolved data. To demonstrate this, we selected an extended XAS spectrum measured on a NiO thin film sample at the Ni
L$_{3,2}$ edge. The sample was excited above its band gap by an optical laser pulse
with 266~nm wavelength, 50~fs pulse duration, and 5~mJ/cm$^2$ peak fluence. The time delay between
optical pump and X-ray probe was fixed to 1.0~ps. During the FEL train, 18 pairs
of optical pump X-ray probe were used, with 17.8~$\mu$s separation between them, corresponding
to an effective repetition rate of 56~kHz during the train.

\begin{figure}
\caption{a) XAS and b) transient change in XAS at the Ni L$_{3,2}$ edges of a NiO thin film. The data were recorded
at a fixed time delay of 1.0~ps, with a peak fluence of 5~mJ/cm$^2$ and a pump wavelength of 266~nm.
Data that are only dark-corrected are shown as continuous blue lines and labeled ``raw''. Data that are additionally
flat-field corrected are shown as dashed orange lines and labeled ``flat-field''. Data that are also corrected
for non-linearity are shown as dotted--dash green lines and labeled ``non-linear''.
}
\label{fig:XASdXAS}
\includegraphics[width=\figurewidth]{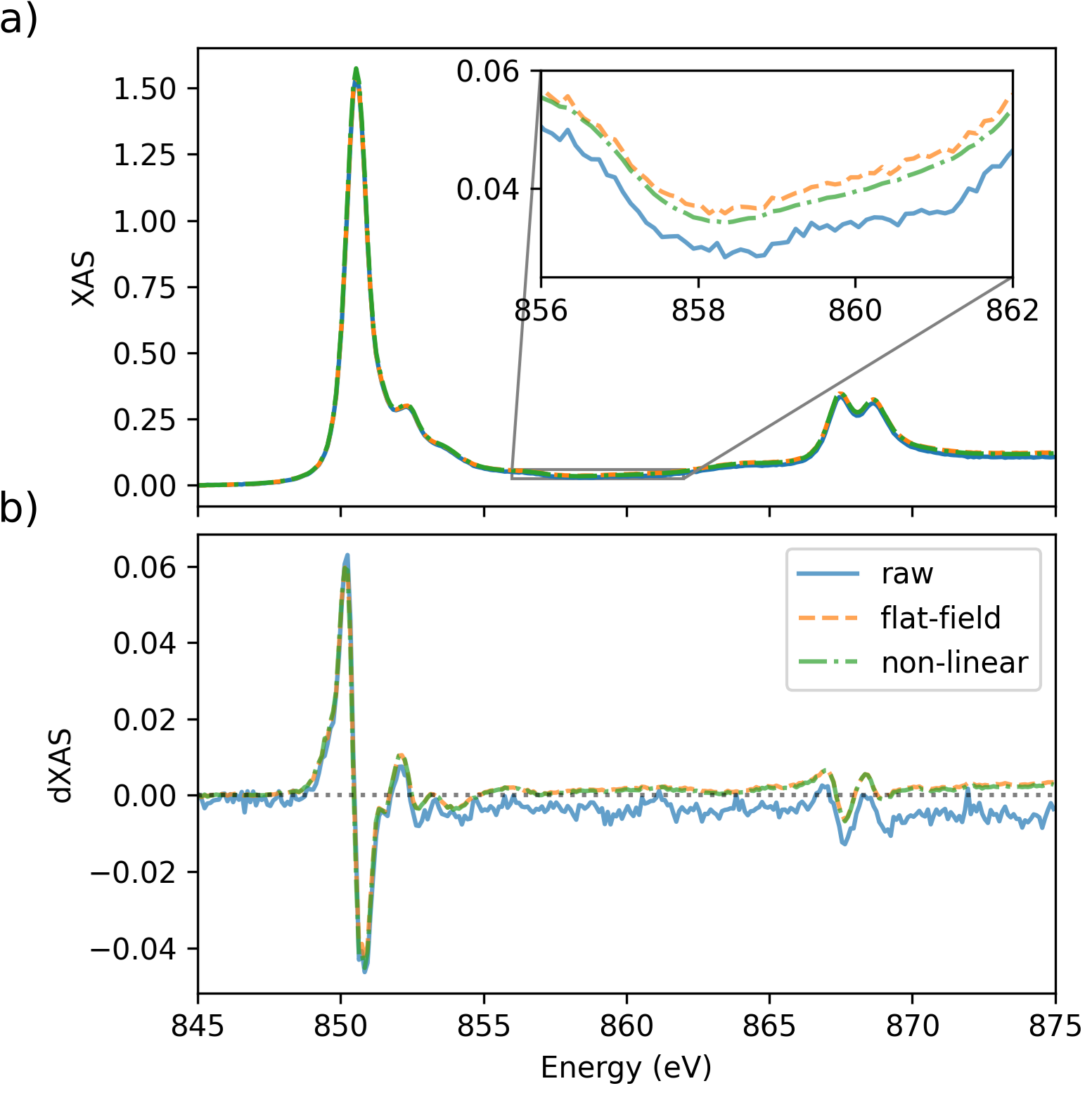}
\end{figure}

In Fig.~\ref{fig:XASdXAS}~a), the XAS and b) the pump-induced change in XAS at the Ni
L$_{3,2}$ edges are shown as continuous blue lines and labeled as ''raw'' for data that are only dark-corrected.
Discussions of the physics behind these NiO XAS and transient XAS are beyond the scope of this article
and will be published separately.
However, we can briefly discuss the XAS and transient XAS measured. The XAS is characteristic
of NiO and its multiplet structure \cite{Regan2001, Groot2021}. In the simplest picture, the transient XAS probes the holes left by
excited electrons after interaction with the pump laser and is observed as an increase in absorption.
At the same time, the states filled by excited electrons give a reduction of the XAS \cite{Stamm2007,
Boeglin2010, Willems2020, Hennes2020, LeGuyader2022}.

The amplitude of the transient XAS is of the order of ~5\% of the static XAS at most, as seen in Fig.~\ref{fig:XASdXAS}~b).
Although small, it is above the noise in the measurement even for the ``raw'' data that are
only dark-corrected. Here we define noise as the apparent
random fluctuations in data points that are close to each other. The situation improves drastically using
the flat-field corrected data,
which are shown as a dashed orange line in Fig.~\ref{fig:XASdXAS}. Over the whole spectrum, the noise level
in the transient XAS is significantly reduced
compared to the uncorrected data. In the XAS spectra, this reduction is also visible in the inset in
Fig.~\ref{fig:XASdXAS}~a), which shows a zoomed region on the flat continuum transitions part of the XAS between
the L$_3$ and L$_2$ edges. We note that for this dataset, the characterization run without sample
in the beam was not recorded at the time. To compute the flat-field and non-linear corrections, we
instead selected in the data the shots falling in the flat pre-edge region below 848~eV. We do not
expect the analysis to be significantly affected by this, as confirmed in experiments conducted later.
The data, that are in addition
corrected for non-linearity, are shown as a dotted--dash green line in Fig.~\ref{fig:XASdXAS}. For
the XAS, a further improvement in the data is visible in the zoomed inset, where the curve is now very smooth
with negligible noise remaining. For the transient change, there is almost no visible difference between
flat-field correction and non-linear correction. This is what one would expect from the results shown
in Fig.~\ref{fig:corr},
where the improvement with the non-linear correction is limited for the -1\ts{st}/1\ts{st} order, which
directly probes the transient XAS change, while the improvement is much larger for the other two ratios
probing the unpumped and pumped XAS spectra.

While it is evident that each additional correction improves the data quality with
a significant noise reduction, there are also some systematic deviations. This is evidenced by the fact
that the three different level of corrections do not result in curves overlapping with each other
in Fig.~\ref{fig:XASdXAS}. Therefore, we have to discuss the implications of each correction on the data. 
The flat-field correction ensures that the measured intensity is
independent of the FEL intensity profile impinging on the zone plate. Clearly,
the transmission of the sample should not depend on the pulse-to-pulse fluctuating FEL intensity
profile, therefore, data that are not corrected by the flat field cannot be trusted. Similarly, the non-linear
correction ensures that the measured quantity does not depend on the X-ray intensity, so data that are
not corrected for the remaining DSSC non-linearity cannot be trusted either. Are there systematic variations
in flat-field and non-linearity corrected data that require additional correction? There is one, which is visible
in Fig.~\ref{fig:XASdXAS}~b), where the baseline of the transient XAS seems to shift away from zero.
This is probably related to the flat-field correction, which is calculated
at a fixed photon energy. However, the zone-plate properties depend on the photon energy. This can be corrected
with an additional step, where we record a XAS spectrum without sample, from which the linear background can be extracted
and subtracted from flat-field and non-linearity corrected data. Such data are not available for this
particular data set, so this correction cannot be applied here. It is however now part of the standard
measuring protocol.

Overall, we have shown that the different data corrections allow extraction of the most information
from every photon detected and to record XAS and transient XAS with excellent SNR, approaching
the photon shot-noise limit.

\subsection{Setup limits\label{sec:limits}}

As we have shown, the BOZ setup at SCS allows recording XAS with sensitivity reaching
the photon shot-noise limit. Detecting more photons or increasing the repetition rate of the experiment
are two simple means, by which we can increase the statistics. In this section, we will thus
discuss the limits in terms of X-ray photons that we can use and the limits in terms of
repetition rate. Finally, given these limits, we will review which sample systems
can be measured with this setup.

\subsubsection{X-ray fluence\label{sec:Xrayfluence}}

If our setup is photon shot-noise limited, then counting more photons by increasing the beam intensity
would directly translate into a better signal with less noise, as confirmed by Fig.~\ref{fig:SNR}.
This is true until we reach saturation of pixels in the DSSC detector too frequently, as such data
have to be discarded from the analysis, reducing thereby the final statistics. In practice,
the X-ray intensity is adjusted such that few percents of the shots are saturated. If the intensity
of the monochromatic X-ray could be made more stable, for explain by employing a self-seeding scheme \cite{Serkez2013},
frequent pixel saturation could be avoided while collecting intense shots more regularly, resulting
in a higher final photon counts.
Given this pixel saturation and the lack of available soft X-ray seeding scheme at SCS,
the only way to increase the intensity would be to enlarge the beam even more on the DSSC sensor.
Here we reach two limits with the current setup. First, expanding the beam even more means that
we need to either move the DSSC further downstream or use a BOZ with a shorter focal length.
However, the DSSC detector is already placed as far downstream as possible given the current size
of the SCS experimental hutch.
Using a BOZ with a shorter focal length would reduce the space available
between the BOZ and the sample to couple in the optical pump laser. Moreover, given that with
the current setup, we are nearly fully illuminating the sensor, as seen in Fig.~\ref{fig:ff},
expanding the beam further would require a larger monolithic sensor. Clearly, given that the
DSSC detector is currently the only detector capable of recording up to 800 pulses per train
at 4.5~MHz repetition rate that can be delivered by the European XFEL at SCS,\cite{Porro2021}
we are at the limit.
We could only make use of higher beam intensity with a new detector having higher saturation limits,
a larger continuous sensor, or smaller pixels with similar electron well depth.

\begin{figure}
\includegraphics[width=\figurewidth]{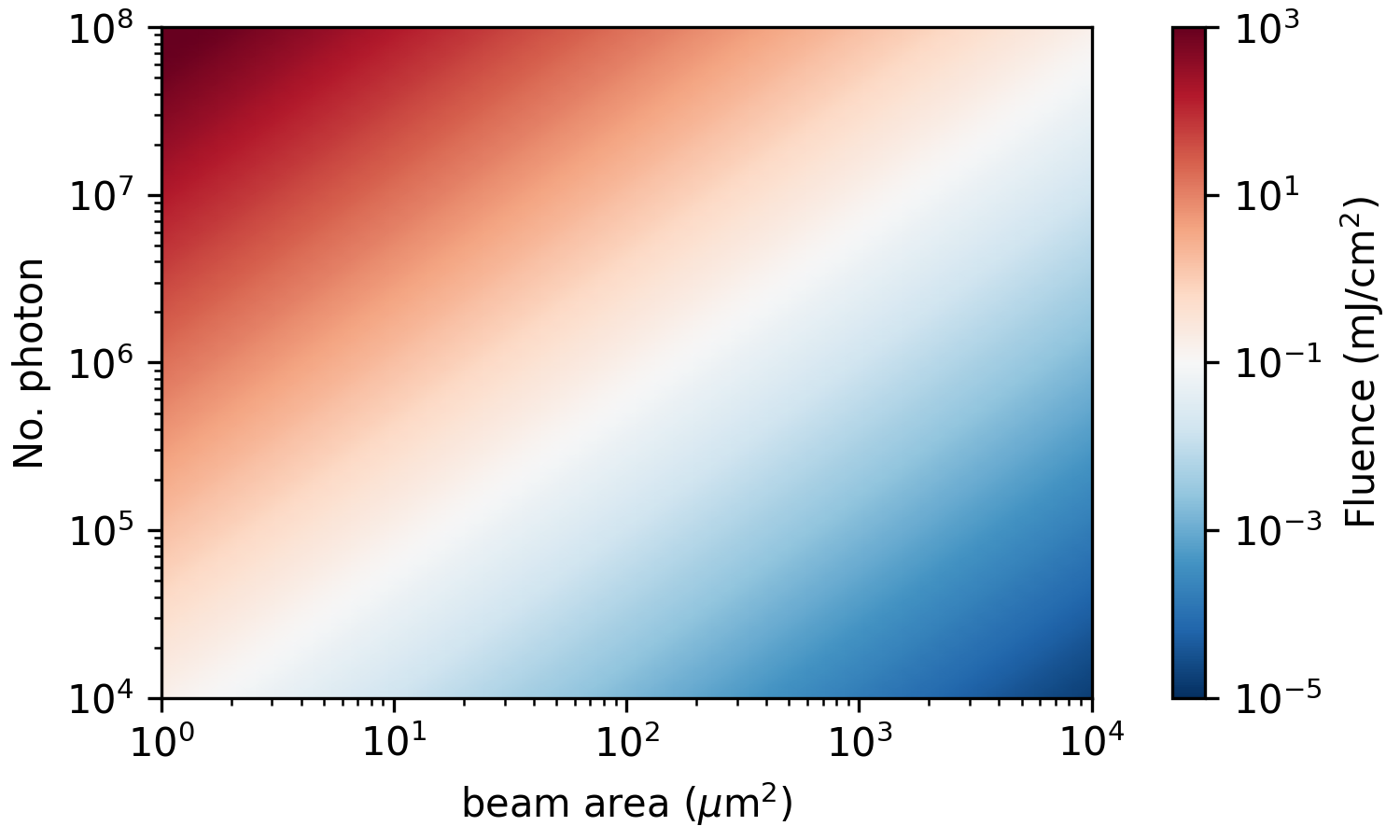}
\caption{X-ray fluence on the sample, in mJ/cm$^2$, as a function
of the beam size and the number of 1~keV photons in the beam. A
limit of 0.1~mJ/cm$^2$, ensuring negligible non-linear X-ray
absorption effects, is shown as white color. Values below and
above this limit are shown in blue and red, respectively.}
\label{fig:BOZ-fluence-limit}
\end{figure}

However, detecting more photons is not the only aspect we should discuss here. With increasing the
X-ray intensity, eventually non-linear X-ray absorption effects will set in, where the X-ray pulse
modifies the sample that it probes \cite{Wu2016, Higley2019}. In Fig.~\ref{fig:BOZ-fluence-limit},
we show the X-ray fluence as a function of the X-ray spot size on the sample and the number of photons
in the beam for photons with 1~keV energy. Non-linear X-ray phenomena set in at fluences of few mJ/cm$^2$ \cite{Wu2016, Higley2019},
so we need to stay below 0.1 mJ/cm$^2$ to be on the safe side. With a typical beam size of 30x30~$\mu$m$^2$
the beam intensity should stay below few 10$^6$ photons. This is already the range we reached, as shown
in Fig.~\ref{fig:corr}, so we are close to the limit here as well. It is possible, in principle, to
increase the X-ray beam size on the sample by simply moving it further downstream of the zone-plate focus. However,
a larger X-ray beam size means an even larger optical pump spot size, which results in slower heat dissipation. As discussed
in the next section, slow heat dissipation limits, in turn, the number of X-ray pulses that can be used per train
such that the beneficial effect of increased statistics per shot with increased spot size might be compensated
by a reduced number of shots due to sample heating.

\subsubsection{Repetition rate\label{sec:reprate}}

In stroboscopic pump-probe experiments, heat dissipation is a known issue limiting the effective
repetition rate at which data can be collected. This is particularly the case here, as the samples
are X-ray transparent thin membranes, which limit the heat dissipation to the two in-plane dimensions.
Moreover, the X-ray pulse pattern at the European XFEL, with its trains of X-ray pulses at
up to 4.5~MHz, leaves very limited time between the X-ray pulses in the train 
for heat dissipation to take place.

\begin{figure}
\includegraphics[width=\figurewidth]{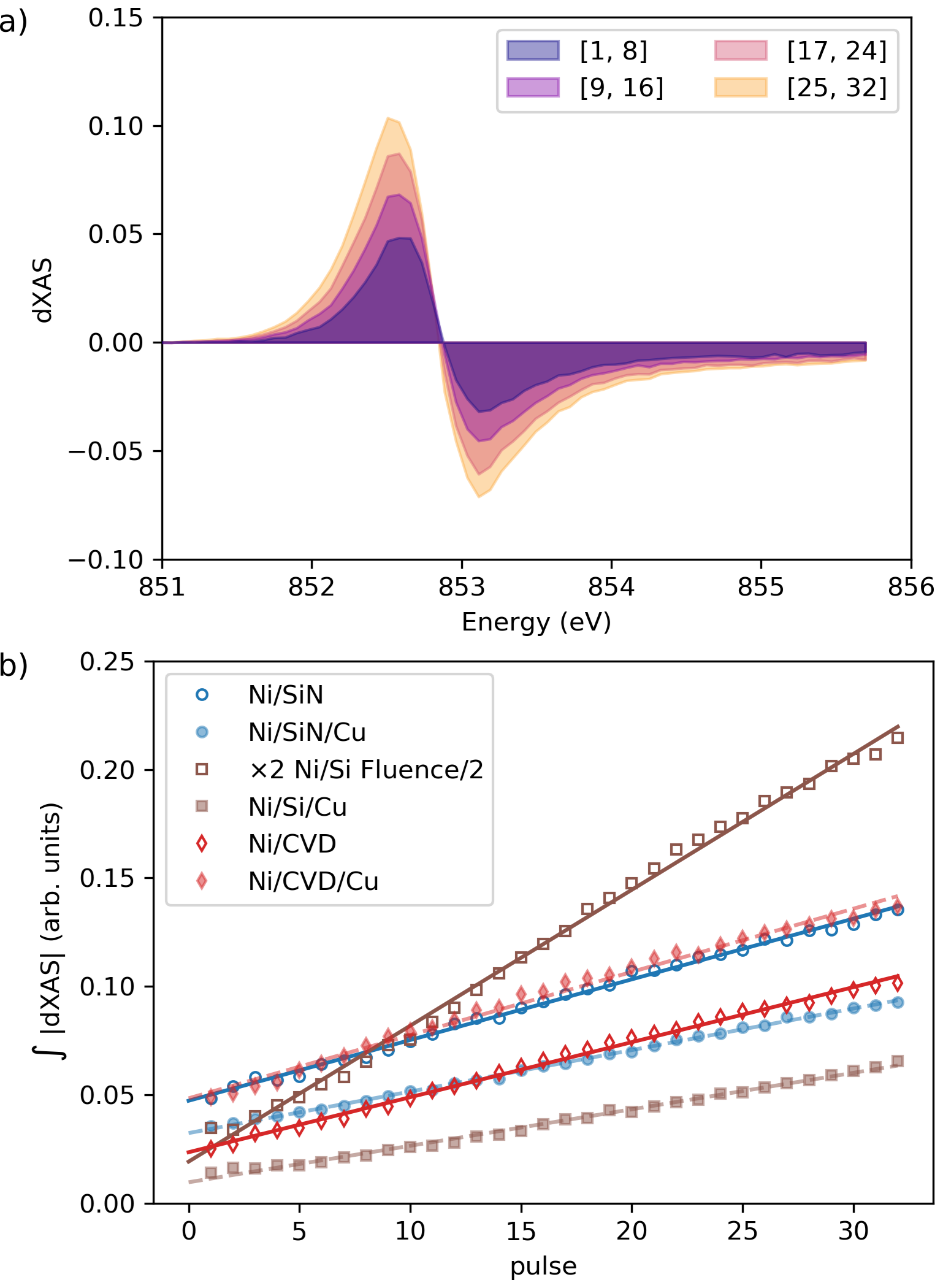}
\caption{a) Transient change in XAS in the Ni/Si$_3$N$_4$ sample as a function of
photon energy for selected pulses in the train as given by the pulse range
in the corresponding legend. The time delay was 0.5~ps, the laser fluence was
7~mJ/cm$^2$, and the pump wavelength was 800~nm. The repetition rate of the 32 pulses
during the FEL train was 280~kHz.
b) Integral of the absolute change in XAS as a function of the pulse number in the train,
ranging from 1 to 32. The points are the measured data, while the lines are linear fits.
}
\label{fig:CWheating}
\end{figure}

\begin{table}[]
    \caption{Slope and intercept and their ratio (intercept/slope)
    fitted from the integral of the absolute change in XAS as a function of the pulse number in
    the train shown in Fig.~\ref{fig:CWheating}. The samples are 20~nm Ni film capped with 2~nm MgO
    and grown on different sample stacks listed below with thickness in nm.}
    \label{tab:CWheating}
    \begin{tabular}{llrrr}
    membrane & heat sink & slope & intercept & ratio \\
     &  & (10$^{-3}$) & (10$^{-2}$) & \\
    \hline
    Si$_3$N$_4$(200) & ---  & 2.8 & 4.7 & 16.9 \\
    Si$_3$N$_4$(200) & Cu(100) & 1.9 & 3.2 & 16.9 \\
    Si(200) & ---  & 6.3 & 1.9 & 3.0 \\
    Si(200) & Cu(100) & 1.7 & 1.0 & 5.7 \\
    CVD diamond(100) & ---  & 2.6 & 2.3 & 9.3 \\
    CVD diamond(100) & Cu(100) & 2.9 & 4.8 & 16.6 \\
    \end{tabular}
\end{table}

In Fig.~\ref{fig:CWheating}~a), the transient change in XAS is shown as a function of the photon
energy around the Ni L$_3$ absorption edge in a 20~nm Ni sample for different ranges of optical pump X-ray probe
pulse pairs within each train.
A clear trend is visible, where the change in XAS increases with the pulse pair number. In a simple
picture, we can interpret this change
in XAS as a change in electron population around the Fermi level. The integral of the absolute change
in XAS over the spectral range measured is then proportional to the deposited energy in the system.
We plot this deposited energy as a function of the pulse pair
number in the train, for different sample stacks, as shown in Fig.~\ref{fig:CWheating}~b). The same Ni
film is deposited on membranes made of silicon, silicon nitride, and diamond, with and
without a Cu heat sink for each case. The diamond membranes are prepared by chemical vapour deposition
(CVD). All data in Fig.~\ref{fig:CWheating} appear to be linear, i.e.
each optical pump pulse in the train adds energy to the system that does not dissipate
completely before the next pulse arrives, leading to a temperature increase of the sample.
We fitted these data with a straight line and extracted
the slope and intercept, as listed in Table.~\ref{tab:CWheating}. In the ideal case, the sample would be
efficiently excited with the first pump pulse, meaning that we would measure a large intercept. At the
same time, the sample would cool down efficiently until the next pump pulse arrives, meaning that we would
measure the same excited sample for subsequent pulses without any heat accumulation, resulting in the
slope being zero. Therefore, an optimal sample stack combines a large intercept and a small slope. To characterize
this property of the sample stack, we introduce a figure of merit as the ratio of the
intercept and the slope. The larger this
number, the better suited the sample stack is for repetitive pump-probe experiments within the train.
Looking at Table~\ref{tab:CWheating}, we can see that the worst sample would be the Si substrate.
Indeed, for this sample stack, we had to reduce the fluence by a factor of 2, otherwise the sample would
break, suggesting that heat accumulation in this sample is strong. For the Si$_3$N$_4$ sample, the
presence of the Cu heat sink does not improve the performance but both are some of the best sample stacks.
For the CVD diamond substrate, the Cu heat sink improves the performance
by a factor of 2 but does not perform better than Si$_3$N$_4$ based stacks.

In all the cases presented here, the heat accumulation is a measurable effect on top of the transient
change in XAS. If no further analysis is possible, we are forced to limit the number of pulses used in
the experiment and separate them as far as possible, given the European XFEL pulse pattern, which in
practice is often 10 to 20 pulses per train. However, if one can assume the response of the system to be
linear both on the transient change and on the heat accumulation change, then the two contributions can
be disentangled and possibly more pulses per train can be used with improved statistics.

Having discussed the limits in terms of X-ray fluence and repetition rate in the previous section, we will
now discuss from which classes of sample systems can we expect a measureable signal level.

\subsubsection{Sensitivity limits\label{sec:sensitivity}}

To make the best use of the setup and the detected photons, the intensity of the three beams on the DSSC
should be similar. This way, the full dynamic range of the DSSC pixels can be used, from the dark pedestal level
up to the saturation level or to the 4.5 MHz digitization cutoff. If the three beams intensities are not well balanced,
the efficiency of the setup will be reduced, as already discussed with eq.~\eqref{eq:shotnoiselimit}.
In the case of a strongly absorbing sample at a resonant edge, the measurement will be limited both 
by saturation of the DSSC detector on the pre-edge region and by low transmitted intensity at the
resonance. In practice, these considerations limit the sample thickness to one or two absorption length at most.

\begin{table}[]
  \caption{Single-shot static and transient XAS signal as well as SNR for different cases. To achieve
  an SNR greater than 3 for the transient XAS signal, the minimal number of shots is given in the
  last column.}
  \label{tab:snr}
  \begin{tabular}{l|rr|rrr}
  \multirow{2}{*}{Sample} & \multicolumn{2}{|c}{Static} & \multicolumn{3}{|c}{Transient} \\
  & signal & SNR & signal & SNR & minimal shots \\
  \hline
  Fe (20 nm) & 1 & 250 & 0.05 & 12 & 1 \\
  Fe monolayer (0.287~nm) & \num{2e-2} & 5 & 10$^{-3}$ & 0.2 & 225 \\
  molecule on surface & 10$^{-3}$ -- 10$^{-4}$ & 0.25 -- 0.025 & \num{5e-5} -- \num{5e-6} & 10$^{-2}$ -- 10$^{-3}$ & 10$^5$ -- 10$^6$ \\
  \end{tabular}
\end{table}

In other cases, one might be interested in systems that are much more diluted. These are the cases considered
in Table~\ref{tab:snr}. We begin with the case of a 20~nm thick Fe film, which corresponds to one absorption
length at the L$_3$ resonance. The static XAS signal is thus 1, as shown in Table~\ref{tab:snr}.
For the noise level, we can estimate it to be similar
to a no sample case, as the change in the number of detected photons will be moderate. This gives us a single-shot SNR of
250. For the other cases that we will now consider, the absorption will be smaller, so the number of detected photons,
and therefore the noise, will be constant. The only thing that will change is the level of signal. For example, for this
20~nm Fe film, we can roughly estimate the transient XAS to be around 5\% of the static XAS. In other words, the signal
reduces by a factor of 20, giving us now a single-shot SNR of 12, as shown in Table~\ref{tab:snr}. Ideally, we would like to measure
the transient XAS with an overall SNR greater than 3, which can be achieved in this case with a single shot, as indicated
in the last column of Table~\ref{tab:snr}.

If we now consider the case of an Fe monolayer, the static XAS signal scales down by a factor of 50. This results
in a single-shot SNR of 5 for the XAS and 0.2 for the transient XAS. To achieve an SNR of 3 on the transient XAS,
we will have to average 225 pulses. This is between 22.5~s of data acquisition if the sample can only be pumped
at 10~Hz and 1~s, if the sample can be pumped with 20 shots during one FEL train.

If we now consider the case of a single layer of molecules containing a single Fe central atom,
we would have ten to hundred times less absorbing atoms than in the Fe monolayer case, which means hundreds
to tens of thousands more shots and longer acquisition required to reach the target SNR, making this kind
of experiment challenging even with many pulses per train.

Finally, we described in this article the case of an homogeneous sample, where both the excited and unexcited sample
membranes can be prepared identical. In practice, samples are often inhomogeneous, such that the transient XAS computed from
the excited and unexcited membrane might not be meaningful. In such cases, the transient XAS can be measured with this setup
using the excited and reference membranes, by alternating pumped and unpumped shots during the FEL train. The benefit of photon
shot-noise limited detection and shot-by-shot normalization remain.

\section{Conclusion}

The beam-splitting off-axis zone plate setup, which is available to users at the Spectroscopy \&
Coherent Scattering instrument at the European X-ray Free Electron Laser, was presented in detail. We showed
that two essential data correction steps are necessary to make the best use of the collected photons: a flat-field normalization
that compensates for the inhomogeneous diffraction efficiency of the diffractive optics employed, and a correction
of the remaining DSSC non-linearity. Remarkably, with these two corrections, the resulting data are shown to be close to the 
photon shot-noise limit. In addition, we reviewed several tools that we provide to the users, namely
a beam propagation calculator to help users design their samples to be compatible with the fixed beam position of this
setup, the complete analysis procedure in the form of a Python package and associated Jupyter notebooks, and finally
the online analysis framework to process live the measured data with all the levels of correction available.
We showed an example of transient XAS in NiO at the Ni L$_{3,2}$ edges with unprecedented data quality,
and discussed the effect of the different corrections. Finally, we reviewed the current limits
of the existing setup in terms of the number of photons on the sample, the repetition rate that thin X-ray
transparent samples can accommodate, and the signal level for increasingly dilute systems. We finally conclude
that the current setup is as good as it can be with the current detector, and that we can measure transient XAS
down to few tens of layers of molecules.

Further improvement in the sensitivity of the setup may come from the implementation of a self-seeding scheme,
or by using an improved detector, such as the DEPFET DSSC, featuring a lower readout noise and higher dynamic range than the
current miniSDD DSSC we used here.
With the installation of the Apple-X helical afterburners in the near future, circular- and
linear-polarization dependent transient XAS experiments, such as X-ray Magnetic Circular Dichroism (XMCD)
and X-ray Magnetic Linear Dichroism (XMLD), will become possible. Plans to adapt this setup to flat
liquid jet experiments to investigate, for example, optically driven
transitions in molecules in solution are ongoing.
Finally, we note that the data processing that we presented provides
a clear path for a data reduction strategy that can be applied on-the-fly,
as the needed data corrections are computationally fast.
This means that this method is compatible with quasi-continuous
operation at several tens to hundred kHz. Such regime will become available with
forthcoming FEL accelerators that are being developed and would be ideal for
experiments on X-ray transparent thin solid samples.











\section{Data availability}

All data are available in the main text or the supplementary materials.
The raw data generated at the European XFEL for the experiment UP2161
used in Fig.~\ref{fig:CWheating} are available at doi:~10.22003/XFEL.EU-DATA-002161-00. For the experiment UP2712
used in Figs.~\ref{fig:ff}, \ref{fig:corr}, \ref{fig:fit} and \ref{fig:SNR},
raw data are available at doi:~10.22003/XFEL.EU-DATA-002712-00. For the experiment
UP2589 used in Figs.~\ref{fig:setup} and \ref{fig:XASdXAS}, raw data are 
available at doi:~10.22003/XFEL.EU-DATA-002589-00.

\ack{Acknowledgements}

We thank Carsten Deiter, Carsten Broers and Alexander Reich for technical support.
This research was partly supported by the Maxwell computational resources operated at Deutsches Elektronen-Synchrotron DESY, Hamburg, Germany.
L.L.G. acknowledges the Volkswagen-Stiftung for the financial support through the Peter-Paul-Ewald Fellowship.
This research was funded in part by the Deutsche Forschungsgemeinschaft (DFG, German Research Foundation) - Project-ID 278162697 - SFB 1242.


\referencelist[article]




\end{document}